\documentclass[twocolumn,english,superscriptaddress,prl,nofootinbib]{revtex4-1}
\usepackage{lmodern}
\usepackage{lmodern}
\usepackage[T1]{fontenc}
\RequirePackage[latin1, utf8]{inputenc}

\usepackage{color}
\usepackage{babel}
\usepackage{amsmath}
\usepackage{amssymb}
\usepackage{graphicx}
\usepackage[unicode=true,pdfusetitle,
bookmarks=true,bookmarksnumbered=false,bookmarksopen=false,
breaklinks=false,pdfborder={0 0 1},backref=false,colorlinks=true]
{hyperref}
\hypersetup{
	pdfborderstyle=,urlcolor=orange,linkcolor=blue,citecolor=red}

\makeatletter

\usepackage{etoolbox}

\DeclareMathOperator\E{\mathbb{E}}
\usepackage{bigints}

\usepackage{algorithm}
\usepackage{algpseudocode}
\usepackage{color}
\usepackage{babel}
\usepackage{units}

\usepackage[normalem]{ulem}
\usepackage{bigints}

\usepackage{comment}

\usepackage{tikz}
\usepackage{color}
\usepackage{dsfont}

\usepackage[cal=boondoxo]{mathalfa}
\usepackage{grffile}

\usepackage[caption=false]{subfig}

\catcode`,\active

\catcode`\,12

\newsavebox{\@brx}
\newcommand{\llangle}[1][]{\savebox{\@brx}{\(\m@th{#1\langle}\)}%
	\mathopen{\copy\@brx\kern-0.5\wd\@brx\usebox{\@brx}}}
\newcommand{\rrangle}[1][]{\savebox{\@brx}{\(\m@th{#1\rangle}\)}%
	\mathclose{\copy\@brx\kern-0.5\wd\@brx\usebox{\@brx}}}

\makeatletter
\def\maketitle{
	\@author@finish
	\title@column\titleblock@produce
	\suppressfloats[t]}
\makeatother

\allowdisplaybreaks

\makeatother

\begin{document}
	\title[]{Impact of dendritic non-linearities on the computational capabilities of neurons}
	
	
	\author{Clarissa Lauditi}
	\affiliation{Department of Applied Math, John A. Paulson School of Engineering and Applied Sciences, Harvard University, 02138 Cambridge, MA, USA}
	
	\author{Enrico M. Malatesta}
	\affiliation{Department of Computing Sciences and Bocconi Institute for Data Science and
		Analytics (BIDSA), Bocconi University, 20136 Milano, Italy}
	
	\author{Fabrizio Pittorino}
	\affiliation{Department of Electronics, Information and Bioengineering, Politecnico di Milano, 20133 Milano, Italy}
	
	\author{Carlo Baldassi}
	\affiliation{Department of Computing Sciences and Bocconi Institute for Data Science and
		Analytics (BIDSA), Bocconi University, 20136 Milano, Italy}
	
	\author{Nicolas Brunel}
	\affiliation{Department of Computing Sciences and Bocconi Institute for Data Science and
		Analytics (BIDSA), Bocconi University, 20136 Milano, Italy}
	\affiliation{Departments of Neurobiology and Physics, Duke University, Durham, North Carolina, United States of America}
	
	\author{Riccardo Zecchina}
	\affiliation{Department of Computing Sciences and Bocconi Institute for Data Science and
		Analytics (BIDSA), Bocconi University, 20136 Milano, Italy}

	\begin{abstract}
		How neurons integrate the myriad synaptic inputs scattered across their dendrites is a fundamental question in neuroscience. Multiple neurophysiological experiments have shown that dendritic non-linearities can have a strong influence on synaptic input integration. 
		These non-linearities have motivated mathematical descriptions of single neuron as a two-layer computational units, which have been shown to increase substantially the computational abilities of neurons, compared to linear dendritic integration. However, current analytical studies are restricted to neurons with unconstrained synaptic weights and unplausible dendritic non-linearities. Here, we introduce a two-layer model with sign-constrained synaptic weights and a biologically plausible form of dendritic non-linearity, and investigate its properties using both statistical physics methods and numerical simulations. We find that the dendritic non-linearity enhances both the number of possible learned input-output associations and the learning velocity. We characterize how capacity and learning speed depend on the implemented non-linearity and the levels of dendritic and somatic inhibition. We calculate analytically the distribution of synaptic weights in networks close to maximal capacity, and find that a large fraction of zero-weight (`silent' or `potential') synapses naturally emerge in neurons with sign-constrained synapses, as a consequence of non-linear dendritic integration. Non-linearly induced sparsity comes with a second central advantage for neuronal information processing, i.e. input and synaptic noise robustness. We test our model on standard real-world benchmark datasets and observe empirically that the non-linearity provides an enhancement in generalization performance,  
		showing that it enables to capture more complex input/output relations. 
	\end{abstract}
	
	\maketitle
	
	Understanding the computational capabilities of single neurons is among the most fundamental open problems in neuroscience. A long-standing question concerns the role of dendrites in shaping neuronal information processing. In the simplest scenario,  dendrites are devices that sum synaptic inputs linearly, propagating a dot product of a vector of pre-synaptic activity with a vector of synaptic weights to the axon initial segment, where a thresholding operation is applied to decide whether the neuron emits an action potential or not. In this view, neurons are analogous to simple perceptrons, whose learning capabilities have been studied extensively (e.g.~\cite{minsky1988,cover65,Gardner_1988,engel-vandenbroek}).
	
	However, this view ignores the presence of active currents in dendrites, which can potentially lead to non-linear integration of synaptic inputs (for reviews, see e.g.~\cite{london_hausser, larkum2013,major2013}). These nonlinearities are due to various types of voltage-gated ionic currents, such as NMDA receptor mediated synaptic currents \cite{schiller00,branco11}, calcium currents \cite{larkum99,larkum04}, or sodium currents \cite{nevian07}. In cortical pyramidal neurons, in particular, it has been shown that inputs to a single dendritic branch sum in a strongly non-linear fashion, while inputs to distinct dendritic branches sum linearly~\cite{polsky2004computational}. 
	These results have led to the idea that neurons could be better described by multi-layer devices than by the standard perceptron model~\cite{poirazi2003,BRUNEL2014149,ujfalussy18,larkum09,deepneuron,PAGKALOS,sardi2017new}.
	
	Given the non-overlapping tree-like morphology of dendrites and the non-linear integration of synaptic inputs pertaining to the same dendritic branch, a natural choice is to model single neurons as a particular type of a two-layer neural network called a tree committee machine. The computational properties of this neural architecture have been extensively studied in the statistical physics literature. Early works from the 1990s~\cite{mitchison1989, Sompolinsky1992} showed that the storage capacity of tree committee machines increases with the size of the hidden layer~$K$. In the case of the sign non-linearity $g(x) \equiv \text{sign} (x)$, the maximal number of random input/output associations that can be learned scales as $P_c=\alpha_c(K) N$ where $\alpha_c(K)\propto \sqrt{\ln K}$, and~$N$ is the number of inputs~\cite{monasson1995}.
	Recently, it was pointed out that these results are valid only for activation functions presenting a discontinuity at the origin~\cite{Baldassi_2019,pehlevan2021tree}. In particular, in the case of the Rectified Linear Unit (ReLU) non-linearity, an activation function commonly used in machine learning, the capacity of the tree committee machine remains finite as the size of the hidden layer $K$ goes to infinity~\cite{Baldassi_2019}. Moreover, most of the non-linearities used in machine learning  enhance learning by smoothing the corresponding loss landscape and inducing flatter and more robust minima that are attractive for gradient-based algorithms such as Stochastic Gradient Descent (SGD)~\cite{unreasoanable,baldassi2019shaping,Baldassi_2019,lucibello2021deep,Star}. 
	
	The aforementioned studies on the tree-committee machine from the statistical physics/machine learning community 
	are typically performed without including biological constraints on the excitatory/inhibitory nature of synaptic weights (i.e. constraints on their sign), 
	and without considering the specific dendritic non-linearities that are observed experimentally.
	Given that multi-layer networks are well known to have more powerful representation and generalization capabilities than single layer ones, as mathematically shown in early works on approximation capabilities of multilayer neural networks~\cite{HORNIK}, 
	a natural question is to what extent these sign constraints and the specific dendritic non-linearities observed in cortical neurons affect the computational capabilities of single biological neurons.
	
	Here, we set out to study the computational capabilities of a single neuron model with dendritic branches implementing experimentally observed non-linear integration and with sign-constrained positive synapses modeling excitatory connectivity, while inhibitory inputs are incorporated into dendritic and somatic thresholds. 
	Using both analytical methods from statistical physics and numerical simulations we derive the number of possible stored input-output associations (capacity), training speed, noise robustness and its generalization properties to unseen inputs, and compare these properties to neuronal models with different dendritic non-linearities. We also compute the distribution of synaptic weights at maximal capacity and show that this distribution can reproduce experimentally observed synaptic weight distributions in neocortical pyramidal cells. 
	Interestingly, while in standard perceptrons with constrained synapses, sparse synaptic input connectivity can only be obtained when a robustness parameter is introduced \cite{Brunel2004, Brunel_2016}, dendritic non-linear input integration leads to high sparsity even in the absence of any robustness constraint.
	
	
	\section{\label{sec:model}Single neuron model}

	We consider a single neuron model that transforms $N$ binary synaptic inputs $\xi_i=\{0,1\}^N$ into a binary output $\hat{\sigma}=\{0,1\}$. In the standard perceptron model, the neuronal output is 
	\begin{equation}
		\hat{\sigma} = \Theta\left(\sum_{i=1}^{N} W_i \xi_i -T\right)
		\label{eq:perceptron}
	\end{equation}
	where $\Theta$ is the Heaviside function, $\boldsymbol{W}$ is a vector of synaptic weights, typically optimized by a learning process, and $T$ is a threshold. 
	
	Here, motivated by experiments that have revealed significant non-linearities in the summation of inputs within single dendritic branches, but not across branches \cite{schiller00,polsky2004computational,larkum2013}, we consider a generalization of the perceptron model with $K$ dendritic branches, and non-linear summation of inputs within each dendritic branch (see Fig.~\ref{fig:architecture}). In this model, the neuronal output $\hat{\sigma}$ is
	\begin{subequations}
		\begin{eqnarray}
			\label{eq:neuronoutput}
			\hat{\sigma} & = & \Theta(\Delta) \\
			\label{eq:inputsoma}
			\Delta & = & \frac{1}{\sqrt{K}} \sum_{l=1}^{K}  g(\lambda_{l}) - \sqrt{K} \theta_s \\
			\label{eq:inputdendrites}
			\lambda_l & = & \sqrt{\frac{K}{N}} \sum_{i=1}^{N/K} W_{li} \xi_{li} - \sqrt{\frac{N}{K}} \theta_d
		\end{eqnarray}
		\label{eq:neuron}
	\end{subequations}
	where $\Delta$ is the total input to the soma, proportional to the sum of the outputs of all dendritic branches; $g$~is a non-linear function describing the dendritic non-linearity; $\lambda_l$~is the total input to dendritic branch $l$, which is a linear sum of inputs to this branch $\xi_{li} \in \{0,1\}^{N/K}$, weighted by synaptic efficacies~$W_{li}$; $\theta_s$ is a somatic threshold and $\theta_d$ is a dendritic threshold. Notice that this model corresponds to a feedforward network with a layer of hidden units endowed with a non-linear transfer function, corresponding to the dendritic branches, and a fixed output summation layer.
	
	\subsection*{Constraints on weights, excitation and inhibition} The efficacy of real synapses is constrained by the identity of the pre-synaptic neuron. Synaptic weights are non-negative when the pre-synaptic neuron is excitatory (glutamatergic), while they are non-positive when the pre-synaptic neuron is inhibitory (GABAergic). Here, we consider for simplicity a scenario in which only the excitatory weights are modeled explicitly and are plastic. Inhibitory synapses are assumed not to be affected by learning, and are lumped together in the two thresholds, $\theta_d$ and $\theta_s$, describing inhibitory synapses onto dendritic branches and the perisomatic region, respectively. Thus, all synaptic weights $W_{li}$ in Eq.~(\ref{eq:inputdendrites}) obey the constraint~$W_{li}\ge 0$.

	\subsection*{Dendritic non-linearity}
	
	Experiments in neocortical pyramidal cells have indicated that the dendritic output is roughly linear at low stimulation intensities, and that it then increases in a strongly non-linear fashion beyond a threshold, before saturating~\cite{polsky2004computational}. 
	
\begin{figure}[t]
		\centering
		\includegraphics[width=1.0\linewidth]{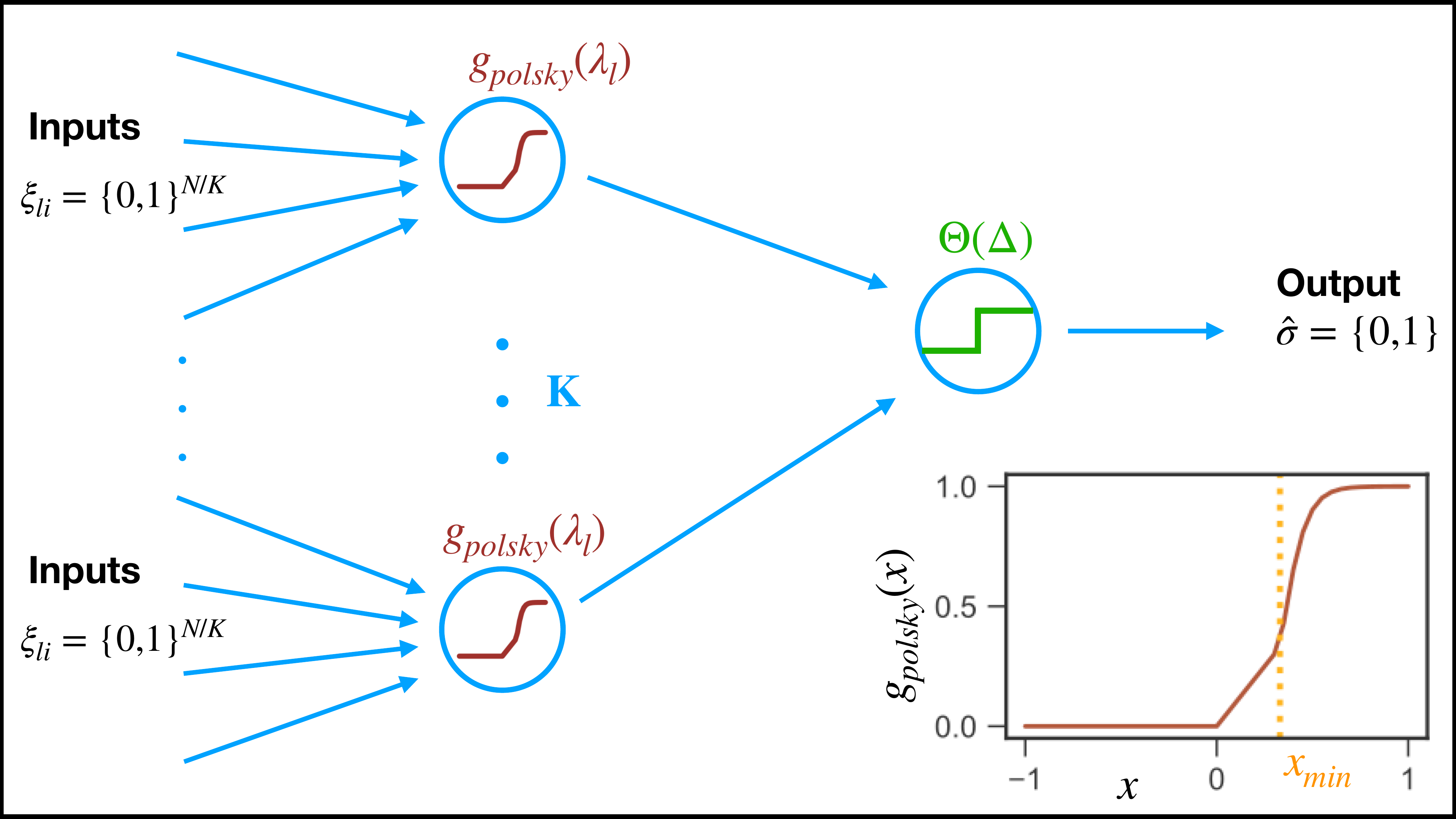}
		\caption{
			Single-neuron model with dendritic non-linearities. The neuron has $K$ dendritic branches as in~\eqref{eq:neuronoutput}. Synaptic inputs to each dendritic branch are summed linearly and then processed through a dendritic non-linearity depicted in the inset. The outputs of the dendritic branches are then summed linearly and compared to a somatic threshold.
		}
		\label{fig:architecture}
	\end{figure}

	To capture quantitatively these findings, we consider the following dendritic non-linear transfer function 
	\begin{equation}
		\label{eq:polsky}
		g_\mathrm{polsky}(x)=\begin{cases} \max(0,x) & x<x_{\min}\\
			\frac{2(1-x_{\min})}{1+e^{-\gamma(x-x_{\min})}}-1+2 \,x_{\min} & x\geq x_{\min}
		\end{cases}
	\end{equation}
	where $x_{\min}$ is a dendritic non-linearity threshold, and $\gamma$ describes the strength of the non-linearity. We refer to this non-linearity as the Polsky transfer function. It is plotted in the inset of Fig.~\ref{fig:architecture}. 
	In the following we use $x_{\min}=0.33$ and $\gamma=15$, which provide a good approximation of the non-linear function measured in~\cite{polsky2004computational}, see Appendix~\ref{sec:bioparams} for a discussion on the biologically realistic values of the parameters.
	Notice that this transfer function interpolates between the ReLU non-linearity (when $x_{\min} \rightarrow \infty$) and the step non-linearity, obtained for $x_{\min}=0$, $\gamma\rightarrow\infty$.
	Note also that in the experiments of ref.~\cite{polsky2004computational}, only excitatory inputs are considered, and consequently only the positive side of the dendritic non-linearity is probed. On the negative side, we take for simplicity $g$ to be equal to zero. This scenario can be thought of capturing in a simplified way shunting inhibition. 
	
	\subsection*{Scaling of inputs and thresholds} 
	Pyramidal cells have on the order of 10,000 synaptic inputs \cite{braitenberg91,iascone20}, scattered along tens to hundreds dendritic branches \cite{elston11}. 
	In this limit, assuming synaptic weights and thresholds are of order $1$ ($W_{li}\sim \theta_d \sim \mathcal{O}(1)$), inputs to dendritic branches scale as $N/K$ due to the sign constraints on the weights, with fluctuations of order $\sqrt{N/K}$ around the mean. To obtain a well-defined limit with finite means and variances, the dendritic threshold should balance the mean inputs, and the difference should be rescaled by $\sqrt{K/N}$. Likewise, at the somatic level, the somatic threshold should cancel the average somatic input, and the difference should be rescaled by $1/\sqrt{K}$. These considerations explain the scalings in Eqs.~(\ref{eq:neuronoutput}-\ref{eq:inputdendrites}). 

	
	\subsection*{Learning tasks} We consider first a standard classification task with the objective of learning a dataset $\mathcal{D} = \{\boldsymbol{\xi}^\mu, \sigma^\mu\}_{\mu=1}^P$ composed of $P=\alpha N$ binary random input patterns $\xi_{li}^{\mu}$ that are i.i.d. Bernoulli variables with $P(\xi_{li}^{\mu}=1)= f_{\text{in}}$ (input coding level) and labels $\sigma^\mu$ that are i.i.d. Bernoulli variables with $P(\sigma^{\mu}=1)= f_{\text{out}}$ (output coding level). The task of the neuron is to correctly classify all input patterns, i.e.~produce the correct output $\hat{\sigma}=\sigma^{\mu}$ when input $\boldsymbol{\xi}^{\mu}$ is presented.
	These input/output associations can be learned by progressively modifying the synaptic weights, either by optimizing directly the number of errors or some surrogate loss functions. This classification task (often called "storage problem" in the literature) has been studied extensively for the perceptron architecture~\cite{Gardner_1988,gardner1988optimal}, including also cases with sign-constrained weights \cite{Amit1989,Brunel2004,Brunel_2016}. It has also been studied in tree committee machines with sign non-linearity on the hidden units~\cite{Sompolinsky1992, monasson1995, stojnic2023capacity}, as well as more recently with generic non-linearity~\cite{Baldassi_2019, pehlevan2021tree}. 
	On the numerical side only, we also study classical benchmark classification tasks in machine learning, providing realistic correlated datasets, such as MNIST \cite{mnist}, Fashion-MNIST \cite{fashion} and CIFAR-10 \cite{cifar10}.
	
	\subsection*{Learning algorithm}
	To evaluate the computational performance of the two-layer non-linear neuron described in~\eqref{eq:neuron}, which is endowed with $K$ dendritic branches and the transfer function defined in~\eqref{eq:polsky}, and compare it with the linear neuron defined in~\eqref{eq:perceptron} from an algorithmic standpoint, we develop an algorithm capable of learning with sign-constrained synapses.  We then proceed to examine its behavior on various paradigmatic learning tasks.
	This algorithm is a modified versions of Stochastic Gradient Descent (SGD)
	as detailed in Appendix~\ref{sec:algoapp} - see Algo.~\ref{alg:SGD}.
	Due to the positive nature of excitatory synapses, whenever the learning rule leads them to become negative, they are instantaneously set to zero. 
	Importantly, the definition of the two models, particularly the tree-like nature of the dendritic layer of the non-linear neuron, naturally allows for their comparison at the same number of synaptic parameters, ensuring that computational improvements are exclusively attributable to their architectural and linear/nonlinear properties.

	\section{Storage capacity}
	
	\subsection*{Analytical methods}
	To investigate the properties of our single neuron model in the storage setting, one can make use of asymptotic methods from statistical physics~\cite{engel-vandenbroek,malatesta2023high}.
	Given a density of patterns $\alpha$, the uniform probability measure over all configurations classifying the patterns in $\mathcal{D}$ (or \emph{solutions} to the learning problem) can be expressed, apart from a normalization factor, as
	\begin{equation}\label{eq::Gibbsm}
		\mathbb{X}_{\mathcal{D}} (\boldsymbol{W}) = \prod_{\mu=1}^P \Theta \left[ (2\sigma^{\mu} - 1) \Delta^{\mu}(\boldsymbol{W};\theta_d,\theta_s) - \kappa \right]
	\end{equation}
	where $\Delta^{\mu}$ is the somatic input defined in~\eqref{eq:inputsoma} and $\sigma^{\mu}$ is the correct label for input $\mu$. The parameter $\kappa$ is a margin that imposes a certain degree of robustness on the learned $\boldsymbol{W}$. Exploiting self-averaging properties, the typical Gibbs entropy, which is the logarithm of the volume of solutions
	can be obtained by taking the average $\langle \cdot \rangle_{\mathcal{D}}$ over the quenched disorder induced by the random realization of patterns and labels
	\begin{equation}
		\label{eq::freeEntropy}
		\phi = \lim\limits_{N, K, P \to \infty}\frac{1}{N} \left\langle \ln \int d\mu(\boldsymbol{W}) \, \mathbb{X}_{\mathcal{D}}(\boldsymbol{W}) \right\rangle_{\mathcal{D}}
	\end{equation}
	In \eqref{eq::freeEntropy}, $\int d\mu (\boldsymbol{W}) \, \bullet \equiv  \int_0^\infty \prod_{li} d W_{li}\, \bullet$ is the integral over the prior weight measure, with the integration bounds reflecting the constraint over the weights. In order to compute the average $\langle \cdot \rangle_{\mathcal{D}}$ in~\eqref{eq::freeEntropy}, one can resort to the Replica Method in the Replica Symmetric (RS) approximation~\cite{mezard1987spin}. We refer to Appendix~\ref{sec:simple_anal} for a brief description of the analytical methods and to Appendix~\ref{sec::Anal} for more detailed derivations.
	
	Pyramidal cells receive roughly 10000 synaptic inputs, which are distributed across several dozen to several hundred dendritic branches \cite{braitenberg91, iascone20,elston11}. We therefore considered the limit of large number $K$ of dendritic branches in our analytical calculations. At the same time, however we consider the regime where $K$ is small compared to the total number of synapses $N$, i.e. $K/N \to 0$. This is not only a realistic assumption, but it also allows us to reduce the computational complexity of the analytical calculations which are valid for a generic non-linearity (see Appendix~\ref{sec:simple_anal}).
	
	As we show in~Appendix~\ref{sec::Anal}, computing the entropy~\eqref{eq::freeEntropy} in the large $N$ and $K$ limit, in turn, gives access to several physical observables of interest, namely the critical capacity and the distribution of synaptic weights.
	
	\subsection*{Critical capacity}
	
	The randomness of the labels in the dataset, does not make the task learnable for any value of the constrained density $\alpha$. Indeed, in the large $N$ limit, there exists a sharp threshold $\alpha_c$ for the probability of finding a solution to the learning problem. For $\alpha <\alpha_c$ this probability is $1$, meaning that there an exponential number of synaptic weight configurations that are able to classify the inputs correctly; at $\alpha = \alpha_c$ the probability of finding a solutions drops abruptly to zero. For $\alpha > \alpha_c$ the complexity of the model is therefore no longer sufficient to classify the activity patterns. $\alpha_c$ can be thought as a measure of expressivity of our single neuron model.
	
	\begin{figure*}[]
		\subfloat[$\kappa=0$, $f_{\text{in}} = f_{\text{out}} = 0.5$ and $\theta_s = 0.5$, $x_{\text{min}} = 0.33$, $\gamma = 15$. \label{fig:alpha_c_vs_thetad}]{\includegraphics[width=1\columnwidth]{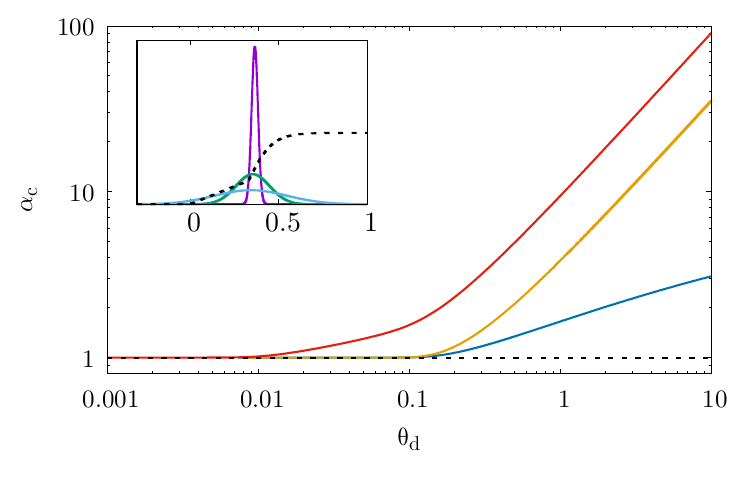}}
		\subfloat[$\kappa=0$, $f_{\text{in}} = f_{\text{out}} = 0.5$ and $\theta_d = 0.5$, $x_{\text{min}} = 0.33$, $\gamma = 15$.\label{fig:alpha_c_vs_thetas}]{\includegraphics[width=1\columnwidth]{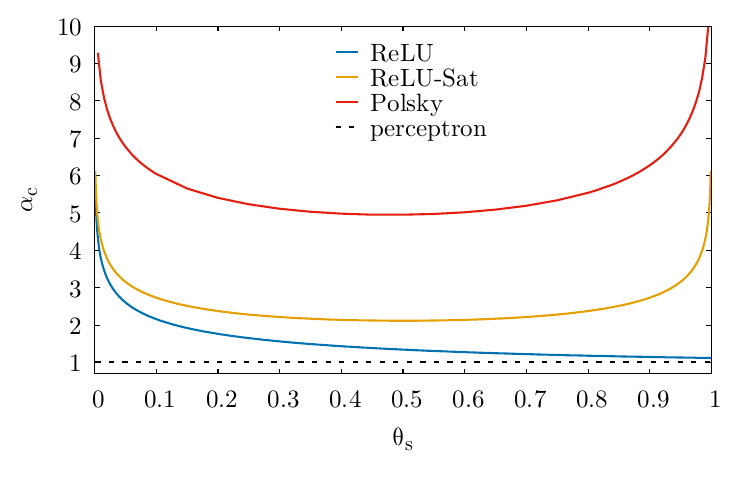} }
		\hspace{1cm}
		\subfloat[$\kappa=0$, $f_{\text{in}} = f_{\text{out}} = 0.5$ and $\theta_s = \theta_d = 0.5$, $\gamma = 15$.\label{fig:alpha_c_vs_xmin}]{\includegraphics[width=1\columnwidth]{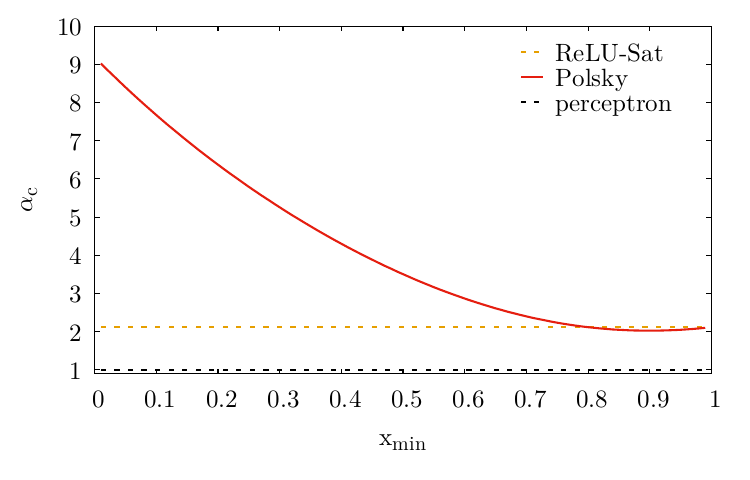} }
		\subfloat[$\kappa=0$, $f_{\text{in}} = f_{\text{out}} = 0.5$ and $\theta_s = \theta_d = 0.5$, $x_{\text{min}} = 0.33$.\label{fig:alpha_c_vs_gamma}]{\includegraphics[width=1\columnwidth]{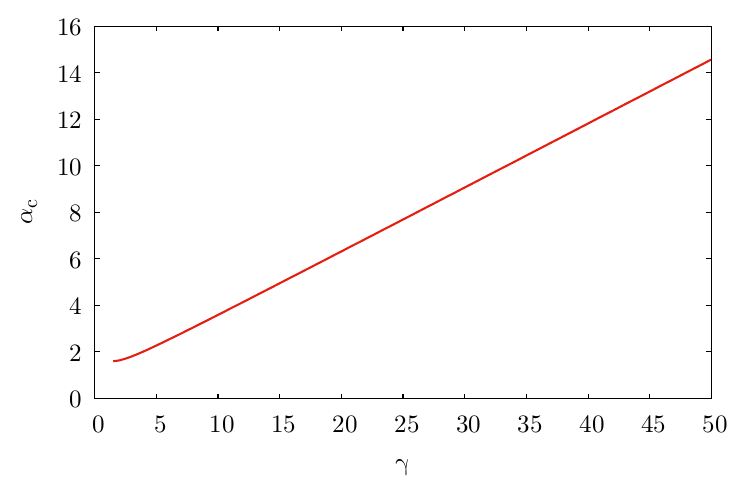} }
		\caption{Critical capacities $\alpha_c$ for ReLU, saturating ReLU and Polsky non-linearities as a function of the dendritic threshold $\theta_d$ (upper left panel), the somatic threshold (upper right panel). The dashed black line represents the case of the one-layer neuron model, where the critical capacity $\alpha_c^{\text{perc}} = 1$. In the inset of the upper left panel we also show the plot of the distribution of the preactivations for $\theta_d = 0.01, 0.05$ and $0.1$ (respectively violet, green and cyan curves); we also plot with the dashed black line the Polsky activation to better show the extent to which the entire non-linearity is exploited for that value of $\theta_d$. In the bottom panels we plot $\alpha_c$ as a function of the parameters of the Polsky activation $x_{\text{min}}$ (left panel) and $\gamma$ (right). In the captions of the panel we show the value of the fixed external parameters.}
		\label{fig:alpha_c}
	\end{figure*}

	At $\alpha_c$ the typical overlap $q$ between pair of solutions extracted from the Gibbs measure~\eqref{eq::Gibbsm} tends towards the typical squared norm $Q$ of solutions. We have therefore expanded the entropy in terms of the variable $dq \equiv Q - q$. We report in~Appendix~\ref{sec::Anal} the technical details of how $\alpha_c$ can be computed from this scaling, for a given value of the external parameters $\theta_d$, $\theta_s$, $f_{\text{in}}$, $f_{\text{out}}$, $\kappa$ and for a generic activation function $g$. 
	
	
	In the case of a linear activation function (i.e. $g(x) \equiv x$), our model is equivalent to the one-layer neuron model whose activity is based on a thresholding operation $\theta_d$ applied to the soma. In this case we recover the results on the critical capacity $\alpha_c^{\text{perc}}$ \cite{Amit1989, Brunel2004, Clopath2012, Brunel_2016}. If the margin $\kappa = 0$ it has been shown in~\cite{Brunel2004} that the capacity is independent on $\theta_d$; in particular, for $f_{\text{in}} = f_{\text{out}} = 0.5$, $\alpha_c^{\text{perc}} = 1$. 
	
	This is not true in the case of the two-layer neuron model, in which changing the dendritic threshold \emph{strongly} alters the expressivity of the model. 
	In the upper left panel of Fig.~\ref{fig:alpha_c} we show the plot of the critical capacity of our two-layer neuron model as function of $\theta_d$ for different types of dendritic non-linearities, namely ReLU, a ``saturating'' ReLU function $\min(\max(0, x), 1)$, and Polsky as in~\eqref{eq:polsky}. For comparison purposes we also plot the critical capacity of the one-layer neuron model. As the figure shows, the storage capacity of the model is greatly enhanced by the presence of the non-linearity.
	As shown analytically in~Appendix~\ref{sec::Anal}, in the limit $\theta_d \to 0$, the two-layer neuron models becomes equivalent to the one-layer perceptron model. When $\theta_d$ increases, all models with non-linear integration increase their capacity, but in a strongly non-linearity-dependent way. With a non-saturating non-linearity such as ReLU, the increase in capacity is logarithmic in $\theta_d$, and it is much smaller than with saturating non-linearities where the capacity increases linearly with $\theta_d$. Finally, the model with Polsky non-linearity outperforms the saturating ReLU function, thanks to the additional non-linear region for $x>x_{\text{min}}$.
	
	The behaviour of $\alpha_c$ as a function of $\theta_d$ can be understood through an analysis of the shape of the distribution of dendritic preactivation, which we show in~Appendix~\ref{sec::Anal} to be a Gaussian with a mean and variance that are functions of the norm of the weights $Q$, and the input coding level $f_{\text{in}}$. We show the shape of the dendritic preactivation for the Polsky activation in the inset of the upper left panel of Fig.~\ref{fig:alpha_c} for several values of $\theta_d$.  If $\theta_d$ is small, the distribution is peaked in a range where the Polsky activation behaves linearly; therefore, the model cannot fully exploit the non-linearity and behaves as a one-layer model. On the contrary, by increasing $\theta_d$, the dendritic preactivation distribution widens towards the region where the Polsky activation saturates; if one keeps increasing $\theta_d$ the weight of the active region before saturation becomes negligible. In this limit, we expect the critical capacity to diverge, since the Polsky activation becomes equivalent to the Heaviside theta activation.
	
	The capacity $\alpha_c$ also strongly depends on the somatic threshold $\theta_s$, as shown in the upper right panel of Fig.~\ref{fig:alpha_c}.
	In the bottom panels of Fig.~\ref{fig:alpha_c} we show how the capacity of the network with a Polsky non-linearity depends on the choice of its parameters $x_{\text{min}}$ and $\gamma$. The critical capacity increases both decreasing $x_{\text{min}}$ or increasing $\gamma$, as in this case the non-linearity is closer to the Heaviside theta function. 
	When $x_{\text{min}}=1$, the Polsky non-linearity effectively reduces to the "saturating" ReLU function, and so the capacity of the two models coincide in this limit.

	\begin{figure*}[t]
		\centering
		\includegraphics[width=0.49\linewidth]{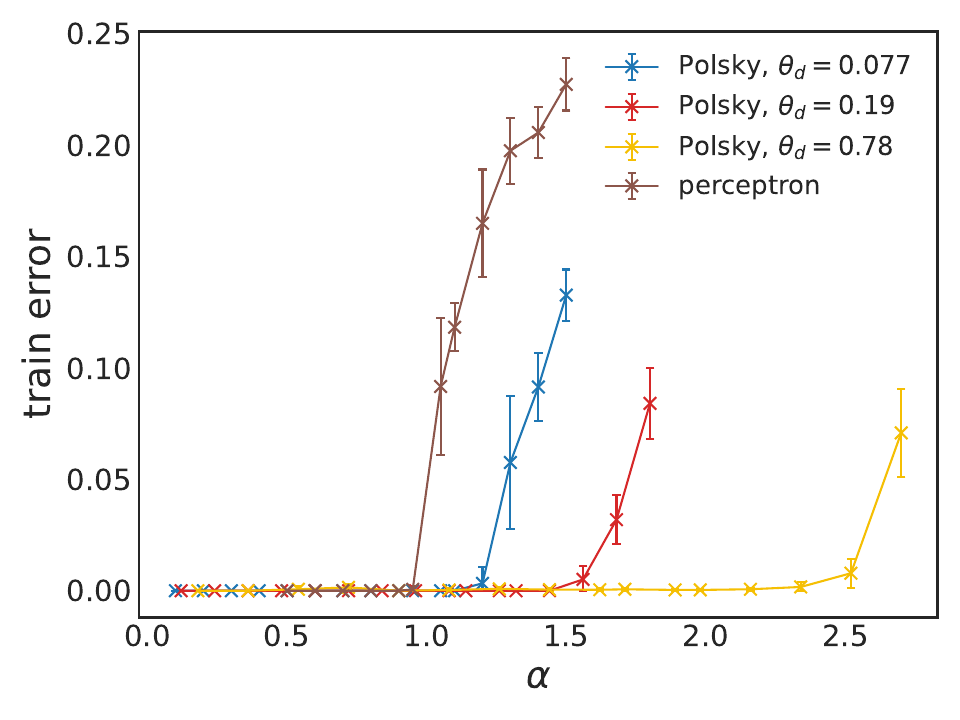}
		\includegraphics[width=0.49\linewidth]{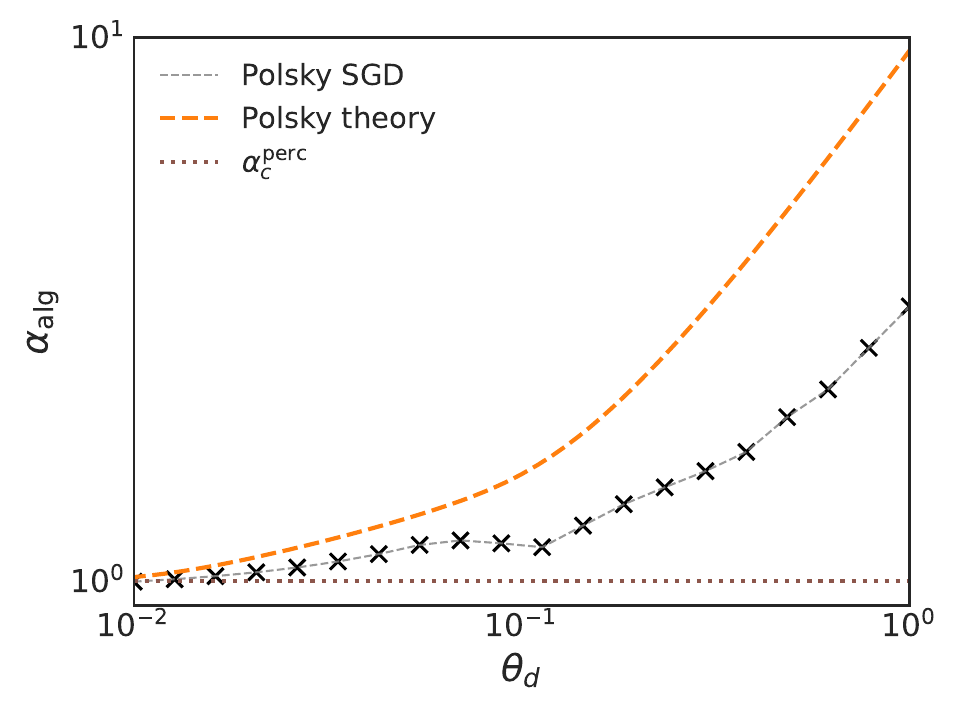}   
		\caption{
			Left panel: \textbf{fraction of misclassified patterns on the training set} as a function of the total fraction of patterns $\alpha=\frac{P}{N}$ for the non-linear neuron compared with the linear one, 
			both trained with SGD (Algo.~\ref{alg:SGD}). The curves show different representative values of the dendritic and somatic thresholds $\theta_d$, with $\theta_s=0.5$. For both neuron models the number of synapses (equivalently, the input size) is $N=999$, the number of dendritic branches for the non-linear neuron is $K=27$, and each curve is averaged over $10$ realizations of the initial conditions. Note that the non-linear neuron achieves capacities greater than the maximal capacity of the linear perceptron model, which is~$\alpha_c^{\mathrm{perc}}=1$. The optimal initial learning rate (see Appendix~\ref{sec:algoapp} for details) for both the linear and non-linear neurons is $\zeta=0.01$. 
			Right panel: \textbf{algorithmic capacity} for 
			SGD~(Algo.~\ref{alg:SGD}), compared to the analytical RS estimate in function of the dendritic threshold~$\theta_d$. Same parameters as the left panel, considering biologically plausible values of $\theta_d \in \left[0.01,\,1\right]$ (see Appendix~\ref{sec:bioparams} for a discussion of the biological values of the parameters).
			The critical capacity of the perceptron (dashed line) is $\alpha_{c}^{\text{perc}}=1$. 
		}
		\label{fig:alpha_alg}
	\end{figure*}
	
	Notice that the estimation of the critical capacity that we have done is based on the Replica Symmetric (RS) ansatz; in general since the model we are analyzing is non-convex, the RS ansatz is thought to be only an upper bound to the true result. In order to get more precise results on the critical capacity, one needs to resort to the Replica Symmetry Breaking ansatz (RSB)~\cite{Parisi1979}. 1RSB corrections to the critical capacity estimation has been computed in one and two layer non-convex neural network models with no constraint on the sign of the weights~\cite{Sompolinsky1992,Engel1992,Baldassi_2019,Baldassi2023Typ,pehlevan2021tree}. Recently, the exact capacity of infinitely wide tree committee machines and perceptrons with negative stability has been computed using a full-RSB ansatz~\cite{fRSB}. For our two-layer neuron model, computing 1RSB effects on the storage capacity is technically very challenging. Note also that our calculations are done in the $K\rightarrow\infty$ limit. Networks with finite $K$ are expected to have a capacity of at most $16\sqrt{\log(K)}/\pi$, the asymptotic behavior of committee machine with step function non-linearity and no constraints on weights~\cite{monasson1995}. For values of $K$ in the range 30-100, this leads to upper bounds in the range 6-7, far below the large $K$ estimates of the capacity shown in Fig.~\ref{fig:alpha_c} for large $\theta_d$. Thus, the benefits of the specific Polsky non-linearity are expected to be the strongest in an intermediate region of values of~$\theta_d$, $x$ and $\gamma$.
	
	\subsection*{Algorithmic capacity and learning speed}
	
	In the previous section, we computed an upper bound for the maximal capacity using a RS calculation. We now turn to the question of the capacity of specific learning algorithms, and to the question of the speed of learning. We use the SGD algorithm (see Appendix~\ref{sec:algoapp} for details of the algorithm).
	It is important to note that, unlike the linear neuron model, the optimization problem in the two-layer non-linear neuron is highly non-convex, and there is no guarantee that algorithms can reach the critical capacity, similar to results concerning binary $\pm 1$ weights models~\cite{locentfirst, unreasoanable, Baldassi_2021_margin}.

\begin{figure}[ht]
		\centering
		\includegraphics[width=1.0\linewidth]{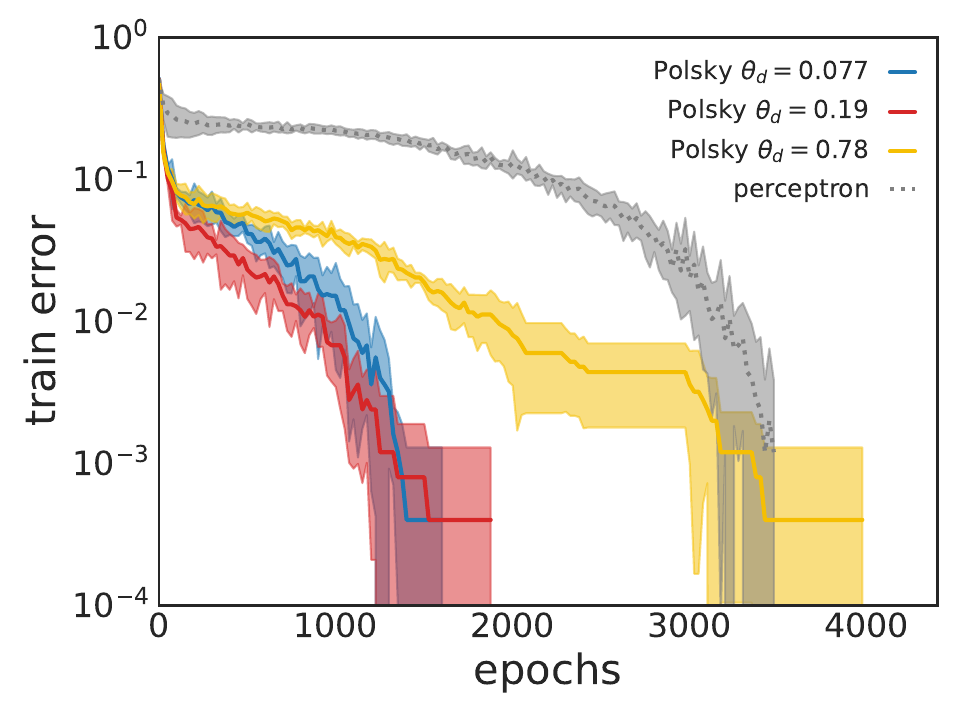}
		\caption{\textbf{Training speed.} 
			Comparison of the convergence times of linear and non-linear neuron models using
			SGD~(Algo.~\ref{alg:SGD}).
			The total fraction of patterns is fixed at $\alpha=0.5$, 
			ensuring that the algorithm can perfectly learn the training set with both the linear and non-linear neuron models.
			For both neuron models, the number of synapses (i.e., the input size) is $N=999$, and the number of dendritic branches for the non-linear neuron is $K=27$. The optimal learning rate for both the linear and non-linear neurons is $\zeta=0.01$. 
			Each curve is averaged over $10$ realizations of the initial conditions.
		}
		\label{fig:speed}
	\end{figure}

	In Fig.~\ref{fig:alpha_alg}, we report the final training error after training with SGD (Algo.~\ref{alg:SGD}) 
	as a function of the control parameter $\alpha=\frac{P}{N}$ (i.e., the density of input patterns) for both the linear and non-linear neuron models. 
	Fig.~\ref{fig:speed} depicts the training error as a function of the training time for SGD. 
	We select the optimal hyper-parameters that maximize the algorithmic capacity and training speed using a grid search procedure (see Appendix~\ref{sec:algoapp}).
	It is worth noting that, at fixed values of the dendritic and somatic thresholds $\theta_d$ and $\theta_s$, the SGD algorithm has two hyper-parameters: the learning rate $\zeta$ and the cross-entropy parameter $\gamma_{ce}$. 
	Appendix~\ref{sec:algoapp} provides an in-depth discussion of algorithmic implementations and hyper-parameter selection for both the algorithmic capacity evaluation and the training speed.
	
	
	As Fig.~\ref{fig:alpha_alg} shows that the neuron with non-linear dendritic integration is able to reach algorithmic capacities that are larger than the maximum one achievable by the linear model, i.e., $\alpha_c^{perc}=1$. 
	Since the comparison between the two neuron models is performed at the same number of parameters (the number of synapses is $N=999$ in both cases), the improvement in performance is solely attributable to dendritic nonlinearities. It is worth noting that in the linear model, the SGD algorithm can reach $\alpha_{alg} = \alpha_c^{perc} = 1$ due to the convexity of the problem.
	Concurrently, as reported in Fig.~\ref{fig:speed}, we find that the non-linear neuron requires fewer steps to learn the training set compared to its linear counterpart.

	Despite the fact that the non-linear neuron algorithmically reaches larger capacities than the maximum capacity theoretically achievable by the linear model, we observe that the algorithmic capacity of our algorithm is generally suboptimal with respect to the analytically calculated critical capacity, $\alpha_c(\theta_d,\theta_s)$ as shown in Fig.~\ref{fig:alpha_alg}. 
	The difference between the SGD algorithmic capacity and the analytical estimate is relatively mild at low $\theta_d$, but the gap widens as the dendritic threshold increases. 
	This difference between algorithmic capacity and analytical RS estimate may be due to several factors, including RSB effects on the critical capacity (as the RS estimate represents an upper bound), finite $K$ effects, and algorithmic hardness. 
	
	\section{Distribution of synaptic weights and input connectivity sparsity}
	
	We next turned to the calculation of the distribution of synaptic weights in our neuron with non-linear dendritic integration. In a perceptron with sign-constrained weights, it has been shown that this distribution contains a delta function at zero (`silent' or `potential' synapses), and a truncated Gaussian distribution of non-negative weights, at maximal capacity \cite{Brunel2004,Brunel_2016},
	\begin{equation}
		\label{eq::P(W)}
		P(W) = p_0 \, \delta(W) + \frac{1}{\sqrt{2\pi} W_\star} e^{-\frac{(W + B W_\star)^2}{2W_\star^2}} \Theta (W) 
	\end{equation}
	where $\Theta(\cdot)$ is the Heaviside function. The fraction of silent weights is a simple function of $B$, $p_0 = H(-B) \equiv \frac{1}{2} \text{Erfc}\left(-\frac{B}{\sqrt{2}}\right)$, whereas $B$, $W_\star$ depend on parameters of the model. In the absence of robustness constraints, the fraction of silent synapses $p_0$ is exactly 50\%, but this fraction increase in the presence of robustness constraints. Furthermore, it has been shown that such a distribution can fit well data from both cerebellar Purkinje cells \cite{isope2002properties,Brunel2004}, and cortical pyramidal cells \cite{song2005highly,Brunel_2016}, but only with a strong robustness constraints, consistent with idea that these networks optimize storage capacity with a strong robustness constraint, or vice versa optimize robustness of stored information. The obtained strong robustness derives from the experimentally observed low connection probabilities, $\sim 0.2$ in granule cell to Purkinje cell connections, and $\sim 0.1$ in layer 5 recurrent pyramidal cell connections.

	It is worth mentioning that, from machine learning standpoint, finding solutions with large margin is desirable for many aspects, such as noise control of input perturbations, or to achieve a good test accuracy. In ~\cite{Baldassi_2021_margin, Baldassi2023Typ} it has been shown that high-margin solutions possess a larger number of flat directions in the loss landscape; this means that one could potentially reduce the number of parameters without a significant performance drop, effectively making the network sparse. As we show in the next section, our two-layer neuron model is able to produce realistic and large connection sparsity by exploiting the dendritic non-linearity, without imposing any explicit robustness constraint.

	\subsection*{Non-linearity automatically induces connection sparsity}

	To investigate the impact of the dendritic non-linearity on both the connectivity sparsity and the whole distribution of synaptic weights, we computed the $P(W)$ of models with dendritic non-linearities. 
	As we show in~Appendix~\ref{sec::Anal}, 
	in the large $K$ limit, the functional form of the distribution of synaptic weights is exactly the same as the one obtained for the perceptron, for any value of $\alpha$, and in particular in the critical capacity limit as in~\eqref{eq::P(W)}. However,  the values of $p_0$, $B$, $W_\star$ depend \emph{strongly} on the type of non-linearity used,
	even when the reliability parameter $\kappa$ vanishes. 
	
	\begin{figure}[ht]
		\centering
		\includegraphics[width=1.0\linewidth]{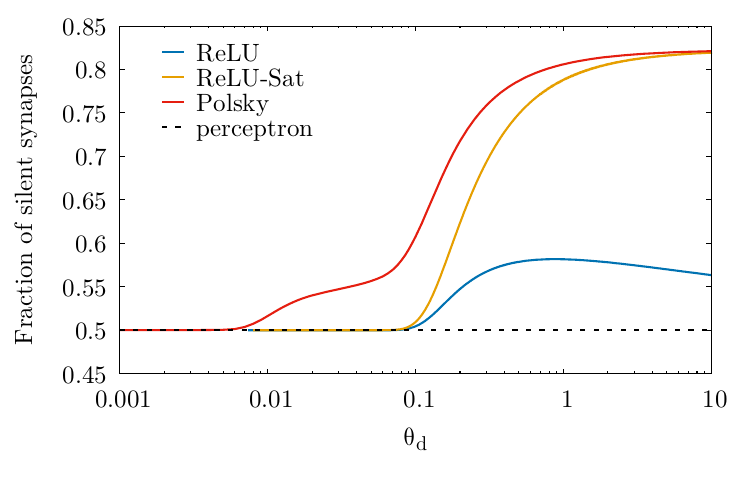}
		\caption{Fraction of silent synapses at maximal capacity as a function of the dendritic inhibition threshold for the Polsky ($x_{\text{min}} = 0.33$ and $\gamma = 15$, red line), ReLU (blue) and ReLU-Sat (orange) non-linearities. Here we have fixed the input and output coding levels $f_{\text{in}} = f_{\text{out}} = 0.5$, the somatic threshold to $\theta_s = 0.5$ and the reliability parameter $\kappa = 0$. The dashed black line corresponds to the case of the linear neuron model, in which the fraction of silent synapses remains constant at 0.5, since $\kappa = 0$.}
		\label{fig:p_0}
	\end{figure}
	
	We show in Fig.~\ref{fig:p_0} the fraction of silent synapses $p_0$ as a function of the dendritic threshold $\theta_d$, for the experimentally measured Polsky and for the ReLU non-linearity at maximal capacity. We see that the experimentally measured non-linearity is capable of greatly increasing synaptic sparsity, in the absence of a robustness requirement during learning. In addition, because Polsky saturates at large preactivations, it is also able to maintain a large value of the sparsity even at large values of $\theta_d$; in this same regime the ReLU non-linearity tends to decrease the number of silent synapses and approaches the sparsity level present in the one-layer neuron model.


	\subsection*{Comparison with experimental data}
	
	We next fitted the analytical distribution of synaptic weights of the non-linear neuron model to the experimental data of $P(W)$ recorded in~\cite{song2005highly} using quadruple patch intracellular recordings of rat cortical pyramidal cells. 
	The observed connection probability in these recordings is $\sim 0.12$. This connection probability is inaccessible to the non-linear model in the absence of robustness constraints. We chose instead  $p_0 = 0.76$, which represents a good  compromise between experimental recordings and biologically plausible parameters that allow us to reproduce $p_0$ with the variables of our model: $\{f_{\text{in/out}}, \theta_{d/s}, \sqrt{N/K}\}$. 
	Once $p_0$ is fixed, we find $W_\star$ (in $mV$) by fitting the experimental histogram of weights with the expression in~\eqref{eq::P(W)}. Choosing $f_{\text{in/out}}=0.05$, we then obtain $\theta_d$ by using the saddle point relation
	\begin{equation}
		\theta_d = f_{\text{in}} W_\star \left[ G(B) - B H(B) \right] 
	\end{equation}
	Finally, we can choose $\theta_s$ in order to obtain from the saddle point equations the value of $p_0 =0.76$ that was chosen initially.
	
	\begin{figure}[ht]
		\centering
		\includegraphics[width=1.0\linewidth]{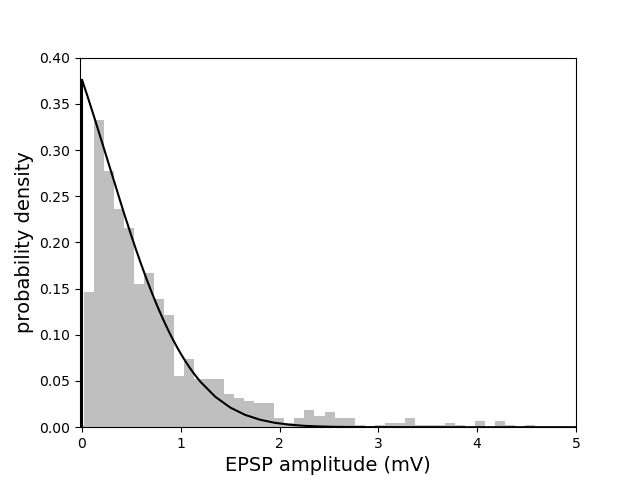}
		\caption{Experimental distribution of synaptic weights vs theoretical distribution at maximal capacity (see equation~\eqref{eq::P(W)}) using as fitting parameters the dendritic and somatic inhibitory thresholds, respectively $\theta_d$ and $\theta_s$. For $f_{\text{in}} = f_{\text{out}} = 0.05$ and $x_{\text{min}} = 0.2$, $\gamma=20$ we have obtained in biological units, we find $\theta_d \simeq 0.6$ mV and $\theta_s \simeq 63$ mV.
		}
		\label{fig:PW}
	\end{figure}
	
	Fig.~\ref{fig:PW} shows that the resulting distribution agrees well with experimental data. We stress that contrary to the case of linear neuron model, our fitting procedure did not rely on the robustness parameter $\kappa$ as a fitting parameter in order to induce sparsity in the model.

	\begin{figure}[ht]
		\centering
		\includegraphics[width=1.0\linewidth]{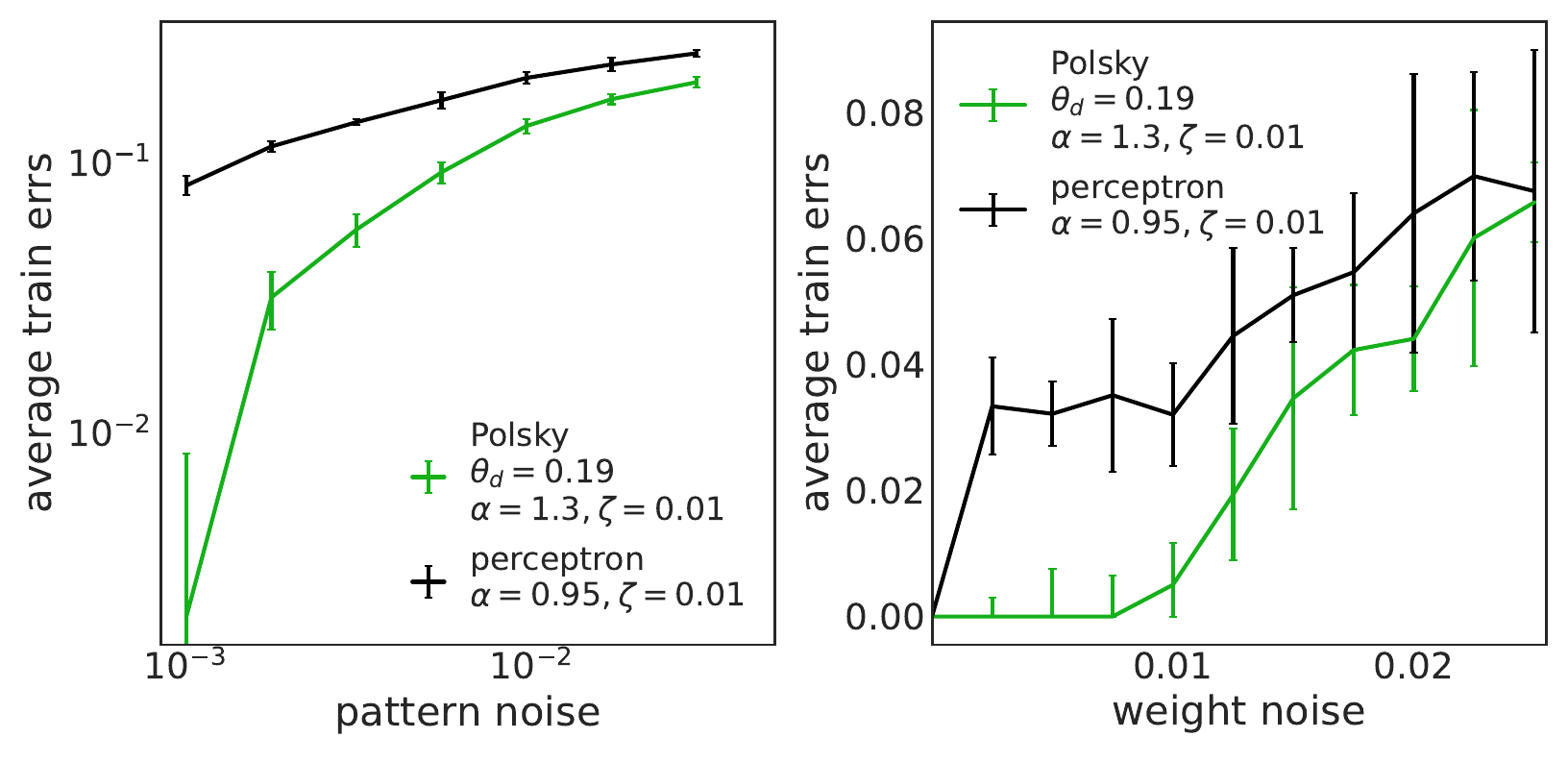}
		\caption{\textbf{
				Robustness to input and synaptic noise in the storage case.}   
			Left panel: Robustness to input noise, measured as the increase in the fraction of misclassified patterns as a function of input flipping probability. 
			Right panel: Robustness to synaptic noise, measured as the increase in the fraction of misclassified patterns as a function of the amplitude of Gaussian synaptic noise. 
			Note that synaptic noise robustness is consistent with the \emph{local energy} definition~\cite{pittorino2021}.
			For both neuron models, the number of synapses (i.e., the input size) is $N=999$, and the number of dendritic branches for the non-linear neuron is $K=27$. 
			Each curve represents the average over $10$ realizations of the initial conditions. 
		}
		\label{fig:weight_distr}
	\end{figure}

	\paragraph*{Synaptic weight distribution at algorithmic capacities}
	Our numerical simulations show that at algorithmic capacity, the distribution becomes well described by a delta function at zero, with a finite fraction of zero weights, and a truncated Gaussian describing positive weights. 
	The fraction of zero weight synapses at algorithmic capacity significantly deviates from the 50\% one corresponding to the linear neuron, although reaching a maximum of~70\% for $\theta_d \gtrsim 1$.
	\begin{figure*}[t]
		\centering
		\includegraphics[width=0.32\linewidth]{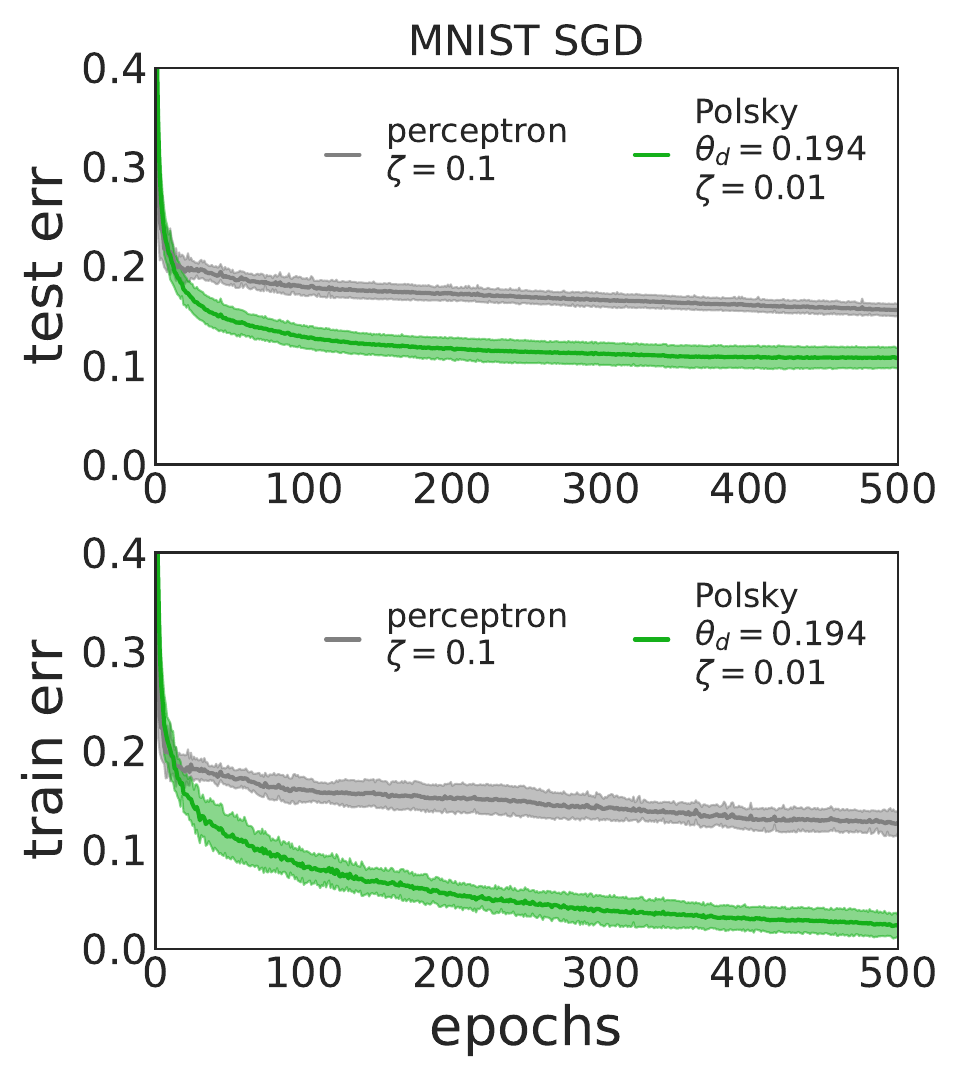} 
		\includegraphics[width=0.32\linewidth]{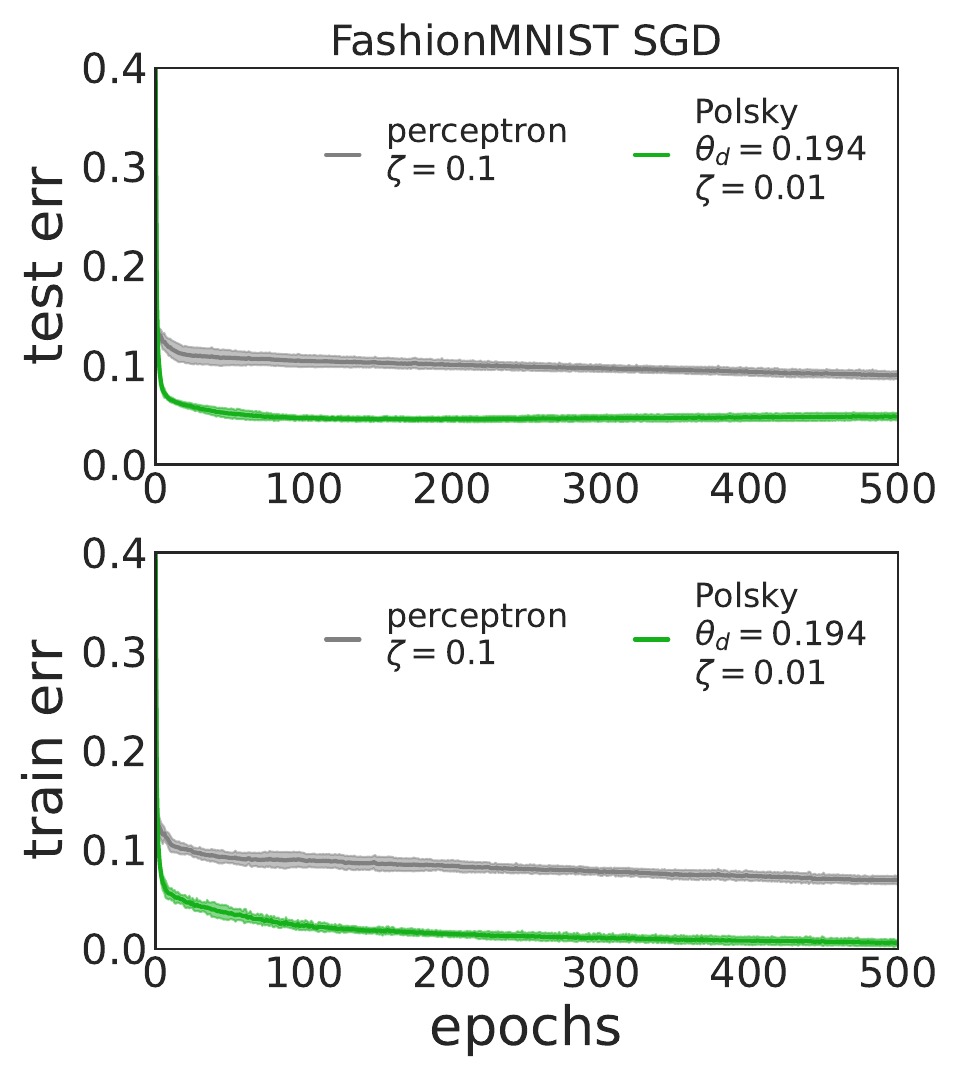} 
		\includegraphics[width=0.32\linewidth]{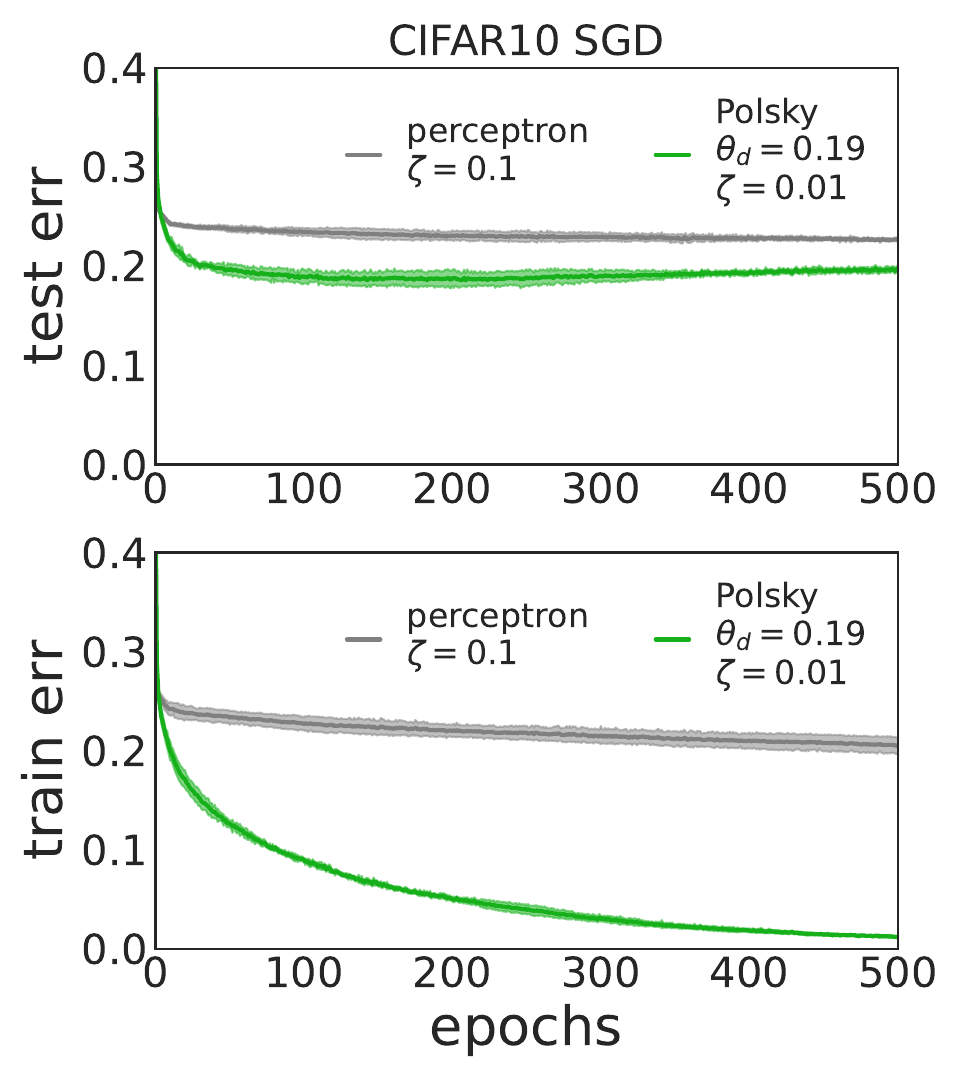} 
		\caption{
			\textbf{Generalization capabilities on real-world datasets: MNIST, Fashion-MNIST, CIFAR-10.} 
			Comparison of train and generalization errors of the non-linear neuron and the linear neuron using SGD (Algo.\ref{alg:SGD}). 
			Left column: MNIST (test and train error in the upper and lower panel respectively); Middle column: Fashion-MNIST (test and train error in the upper and lower panel respectively); Right column: CIFAR-10 (test and train error in the upper and lower panel respectively).
			For both neuron models, the number of synapses (i.e., the input size) is $N=1568$ for MNIST and Fashion-MNIST, and $N=6144$ for CIFAR-10. The number of dendritic branches for the non-linear neuron is $K=49$ for MNIST and Fashion-MNIST, and $K=72$ for CIFAR-10. Each curve represents the average over $10$ realizations of the initial conditions.
		}
		\label{fig:generalization}
	\end{figure*}
	
	\section{Noise robustness and generalization}
	
	\subsection*{Robustness to input and synaptic noise}
	
	Noise is a ubiquitous feature at all levels of the nervous system, from the molecular to the whole brain level~\cite{faisal09}. In particular, single neurons operate in a highly noisy environment, due to background inputs they constantly receive. Thus, robustness to input and synaptic noise are fundamental computational requirements for any realistic single neuron model. We thus turn to an investigation of robustness of our model to noise.  
	In our simulations, we estimate the robustness to input noise by independently flipping the entries of each pattern in the training set with probability $\rho$ maintaining the same label and measuring the train error of optimal synaptic configurations on this corrupted training set. 
	Fig.~\ref{fig:weight_distr} shows the robustness to input noise for the SGD algorithm. We observe that the non-linear model is more robust to input perturbations compared to the linear model with the same number of synaptic parameters.
	Similarly, we measure synaptic noise robustness by estimating the number of misclassified patterns when synaptic strengths are perturbed by applying a multiplicative Gaussian noise of amplitude $\sigma$ to zero-error synaptic configurations $\boldsymbol{W}$. 
	In practice, we measure the quantity $\delta E_\text{train}(\boldsymbol{W}, \sigma) =\mathbb{E}_{\boldsymbol{z}}\, E_\text{train}(\left[ \boldsymbol{W} + \sigma \boldsymbol{z} \odot \boldsymbol{W} \right]_{+}) - E_{\text{train}}(\boldsymbol{W})$, where $E_{\text{train}}(\boldsymbol{W})$ is the number of errors made by configuration~$\boldsymbol{W}$ on the training set, the expectation $\mathbb{E}_{\boldsymbol{z}}$ is over normally distributed synaptic noise realizations $\boldsymbol{z}\sim \mathcal{N}(0, I_N)$,~$\odot$ is the element-wise product, and $\left[\cdot \right]_+$ is the ReLU function. 
	The quantity $\delta E_\text{train}(\boldsymbol{W}, \sigma)$ is also known as the \emph{local energy} in the machine and deep learning literature. It serves as a proxy for the flatness of the energy landscape around a given optimal configuration~\cite{pittorino2021, pittorino22a}. It is worth noting that a synaptic hard threshold at zero is implemented in this case as well, meaning that synaptic configurations cannot take negative values even under perturbations, i.e. $\boldsymbol{W}>0$ and~$\boldsymbol{W}+\sigma \boldsymbol{z} \odot \boldsymbol{W} > 0$). 
	Fig.~\ref{fig:weight_distr} 
	presents the robustness to synaptic noise of the linear and non-linear neurons for the SGD algorithm. The results demonstrate that the dendritic non-linearity enhances synaptic noise robustness, or equivalently, the local energy landscape around optimal synaptic configurations is flatter in the non-linear case.
	
	


	\subsection*{Generalization performance on real-world datasets}
	
	To study the generalization properties of the neuron model defined in \eqref{eq:neuron}, we focus on binary classification learning tasks using the MNIST~\cite{mnist}, Fashion-MNIST~\cite{fashion}, and CIFAR-10~\cite{cifar10} datasets, which are standard benchmarks in machine learning. The generalization error, a fundamental machine learning observable, can only be estimated in the presence of a test set, which is absent in the storage case.
	To ensure that the generalization tasks remain reasonably difficult while still allowing for generalization, we divide the MNIST and Fashion-MNIST datasets into an odd/even binary classification task. For MNIST, we separate odd and even digits into two different classes. Similarly, for FashionMNIST, we separate the classes corresponding to even and odd labels into two groups.  For the CIFAR-10 dataset, we choose two different classes in order to define a reasonably difficult generalization task, namely Bird and Ship.
	Appendix~\ref{sec:algoapp} provides details on the dataset binarization procedure and hyperparameter selection.
	As shown in Fig.~\ref{fig:generalization}, the non-linear neuron demonstrates better performance on all the aforementioned generalization tasks.
	We have also verified that in more challenging scenarios, such as learning odd/even classes in the CIFAR-10 dataset (separating the classes into two groups based on their even or odd labels), the generalization error is very close to random guessing (around $40\%$). 

	\section{Conclusions and future directions}
	
	In this paper, we have studied the effect of realistic dendritic non-linearities on the computational abilities of a single neuron model with sign constrained synaptic weights.
	We have shown that dendritic non-linear integration is beneficial for multiple reasons. Firstly, it enhances the overall expressivity of a single neuron, measured as the maximum number of input-output associations that it can correctly store. Secondly, the non-linearity generates input connectivity sparsity in the model, i.e. it leads to a large fraction of zero weight (silent or potential) synapses, in the absence of any explicit robustness constraint. This is in marked contrast with previously analyzed neuron models with passive dendritic integration, in which a large level of sparsity as recorded experimentally in cerebellar Purkinje and pyramidal cells could only be obtained with a high reliability margin. In addition, the distribution of synaptic weights of our model shows a good agreement with the one recorded in experiments on pyramidal cells. 
	On the algorithmic side, we quantified the benefit of nonlinear dendritic integration from several points of view. Firstly we have shown that nonlinear dendritic processing enables an algorithm such as SGD to find optimal synaptic configurations at larger density of input patterns with respect to linear integration, and to find them consistently faster. Secondly, we have found that synaptic configurations found by such algorithms have desirable computational properties in the non-linear case, such as  a stronger robustness to input and synaptic noise, and higher generalization ability, compared to the linear case.

	Algorithmically reaching the analytically calculated critical capacity $\alpha_c$ is challenging 
	as shown in Fig.~\ref{fig:alpha_alg}. Several factors contribute to the discrepancy between algorithmic and critical capacities. The critical capacity computed using RS serves as an upper bound for the true critical capacity, as RSB is likely to occur at lower values of $\alpha$, a phenomenon well-established for both ReLU and step function non-linearities. Algorithmically, there is no guarantee to reach the optimal capacity, and finite-size effects from simulations with finite $K$ and $N$ could also influence results. Understanding this discrepancy will be the subject of future work.

	Future work will concentrate on dropping some of the modelling assumptions we made here. For example, we have not considered the fact that there may be more than two layers of input processing through dendrites; this would make neurons more similar to deep tree networks. It would be therefore interesting to study the effect of multiple layers of dendritic integration on all the quantities studied in this paper. Similarly, it will be important to describe in a more realistic fashion inhibitory inputs, and the potential impact of inhibitory plasticity. Finally one should also take into account the effects of potential correlations structures in inputs, e.g.~that inputs coming in the same branch are more correlated than inputs pertaining to different branches~\cite{kirchner22}. Similarly, synaptic inputs coming at different regions of the dendritic tree (e.g.~basal vs apical) may have different statistical properties and convey different types of inputs. 
	
	Another important future direction concerns synaptic plasticity algorithms. Here, we have investigated a simple plasticity algorithm (SGD) in a standard supervised learning scenario, in which a teaching signal is available to the neuron. In cerebellar Purkinje cells, this teaching signal could be implemented by the climbing fiber input, which has been shown to correlate with error in motor tasks. In cortical pyramidal cells, the presence of error signals is more speculative. In the presence of the error signal, SGD in a single neuron model leads to a local plasticity rule, that could be plausibly implemented in a biological neuron, unlike in multi-layer networks (see Appendix~\ref{sec:algoapp}). Finally, it will be interesting to investigate the computational abilities of a neuron with non-linear dendrites in a purely unsupervised learning setting.
	
	\section*{Acknowledgements}
	
	This paper is dedicated to the memory of our colleague and friend Luca Trevisan. 
	F.P. acknowledges support by the PNRR-PE-AI FAIR project funded by the NextGeneration EU program. E.M.M. acknowledges the MUR-Prin 2022 funding Prot. 20229T9EAT, financed by the European Union
	(Next Generation EU).

	\appendix
	
	\section{Analytical Methods}
	\label{sec:simple_anal}
	
	The entropy in~\eqref{eq::freeEntropy} can be computed using the replica method; since the average over the log in~\eqref{eq::freeEntropy} is difficult to compute one uses the identity $\ln (x) = \lim_{n \to 0} \frac{x^n -1}{n}$. Taking $n$ as an positive integer one ends to deal with an enlarged system of $n$ identical virtual copies of the system, so that the average can be easily be performed, at the price to couple the replicas together. 
	However in the large $N$ limit, it turns out that the properties of the model can be fully characterized by a finite set of quantities called order parameters that are determined self-consistently by solving equations obtained by saddle point method (see also the recent notes~\cite{malatesta2023high} for a more pedagogical introduction). In practice, the averaged entropy can be fully characterized by the order parameters
	\begin{subequations}\label{eq:orderparams}
		\begin{align}
			&\sum\limits_{i=1}^{N/K} W^{a}_{li} = \frac{N}{K}\overline{W}+\sqrt{\frac{N}{K}}M^{a}_{l}\label{eq:M}\\
			&q^{ab}_{l} \equiv \frac{K}{N}\sum\limits_{i=1}^N W^{a}_{li}W^{b}_{li} \label{eq:qab}\\
			&Q^{a}_{l} \equiv \frac{K}{N} \sum\limits_{i=1}^N\left(W^{a}_{li}\right)^{2} \label{eq:Qab}
		\end{align}
	\end{subequations}
	and their conjugated ones $\hat{M}^a_l$, $\hat{q}_l^{ab}$, $\hat{Q}_l^a$, with $a, b \in [n]$ and $l \in [K]$. They represent respectively: the typical average synaptic weight, the most probable overlap between two replicas extracted from the Gibbs measure in~\eqref{eq::Gibbsm}, and the typical averaged squared norm of a synaptic weights belonging to the dendritic branch $l \in [K]$. Since the fields are non-overlapping, each branch has access only to a portion of the synaptic input; therefore there is no correlation between hidden units in the same architecture. Notice also that, in~\eqref{eq:M}, the average can be expressed with the sum of two contributions: the first represents the average of the scaled synaptic weights~$\overline{W} = \frac{\theta_d}{f_{\text{in}}}$, while the second represents a $\sqrt{\frac{N}{K}}$ correction needed to fine tune this average with respect to the threshold $\theta_d$. We report in Appendix~\ref{sec::Anal} the full analytical calculations of the entropy~\eqref{eq::freeEntropy} in the case in which the structure of the overlap matrix $q^{ab}_l$ is symmetric under permutation over the replica indices (the so called ``Replica Symmetric'' or RS ansatz), and the order parameters do not depend on replica and dendritic branch indices $l \in [K]$. This means $q_l^{ab} = Q \delta_{ab} + q (1-\delta_{ab})$, $\forall a, b \in [n]$ and $l \in [K]$ and similarly for the other order parameters. The entropy can be therefore obtained by maximinizing a function $\phi_{RS}$ with respect to the order parameters 
	\begin{equation*}
		\phi = \max_{q,\hat{q},Q,\hat{Q},M,\hat{M}} \phi_{RS} (q,\hat{q},Q,\hat{Q},M,\hat{M})
	\end{equation*}
	$\phi_{RS}$ can be written as
	\begin{equation}
		\phi_{RS} = \mathcal{G}_{S} + \alpha \,\mathcal{G}_{E}.
	\end{equation}
	i.e. as a sum of an entropic contribution $\mathcal{G}_{S}$ which represent the log of the total volume of configurations $\boldsymbol{W}$, and an energetic part $\mathcal{G}_{E}$ that corresponds to the log of the fraction of solutions for a given $\alpha$. The explicit expressions of $\mathcal{G}_{S}$ and $\mathcal{G}_{E}$ are reported in Appendix~\ref{sec::Anal}.
	
	The values of the order parameters $q,\hat{q},Q,\hat{Q},M,\hat{M}$ can be found by solving a set of coupled saddle point equations obtained by equating to zero the derivative of $\phi_{RS}$ with respect to each of them.

	\subsection{Limit of large number of dendritic branches}
	
	As noted in~\cite{braitenberg91, iascone20,elston11}, pyramidal cells receive roughly 10000 synaptic inputs, which are dispersed across a wide range of dendritic branches, varying from several dozen to several hundred. This motivates considering the limits where $N$ and $K$ tend towards infinity, but with the $K/N$ approaching zero.
	
	On the technical side, solving the saddle point equations for generic $K$ is in general a very difficult task, since in order to compute the energetic term one should evaluate $2K$-dimensional integrals. However in the large $K$ limit the energetic term simplifies considerably. Indeed, because of the Central Limit Theorem, for $K$ large the total inputs to the soma are Gaussian distributed variables with mean and variance that depend on the transfer function implemented by dendrites.
	
	Interestingly, as first observed in~\cite{Baldassi_2019}, the final expression which we derive in detail in~Appendix~\ref{sec::Anal}, becomes equal to an \emph{effective perceptron}, i.e. the entropy is in form equal to the one presented in~\cite{Brunel2004,Brunel_2016} for the one-layer neuron model, but where each order parameter is substituted by an integral expression that depends on the activation function $g$ used.

	\section{Biological parameters}
	\label{sec:bioparams}
	In order to estimate biologically plausible parameters for the non-linear neuron model and the Polsky transfer function, we refer to Fig.~4c of~\cite{polsky2004computational}, for a pair of EPSPs and a single EPSP. In particular, with the saturation of the Polsky transfer function in $x=1,y=1$ on both the $x$-axis and the $y$-axis, we obtain that the biologically realistic ranges for the Polsky parameters are (in units of $15 \, mV$): $x_{\text{min}} \in \left[0.2,0.33\right]$, $\gamma \in \left[13,20\right]$.

	Here below we also report the conversion factors that we have used to convert the quantities  of the model to biological ones. Let's call $W_b$ the biological measurement of a weight (in $mV$) and the corresponding one of the model $W$. They are related by
	\begin{equation}
		W_b = \frac{15 W}{\sqrt{N}} \,.
	\end{equation}
	The dendritic threshold of the model defined in~\eqref{eq:neuronoutput} is related to the biological one by
	\begin{equation}
		\theta_{d b} = \frac{15}{\sqrt{N}} \frac{N}{K} \theta_d = \frac{15 \sqrt{N}}{K} \theta_d
	\end{equation}
	For the fitting procedure of the theoretical distribution~\eqref{eq::P(W)} on the experimental distribution of synaptic weights, we got $W_\star \simeq 5.56 \, mV$, using $p_0 =0.76$. This corresponds, applying the conversions above, to a dendritic threshold of $\theta_d \simeq 0.6 \, mV$, having chosen $\sqrt{N} \simeq K$ and $\sqrt{N} = 100$. The somatic conversion factor is instead
	\begin{equation}
		\theta_{sb} = 15 K \theta_s \,.
	\end{equation}
	
	In order to select reasonable biological values for $\theta_d$, we assume \( N = 10,000 \) (standard synapse count for cortical neurons), \( W_b = 1\,\mathrm{mV} \), giving \( W \approx 6.7 \) (or a reasonable range \( W \in [2, 10] \)). For dendritic inhibition, we consider \( \theta_{db} \) (average inhibition on a branch) in \( [1, 10]\,\mathrm{mV} \). With \( K = 100 \), this implies \( \theta_d \in [0.01, 1] \).  
	\section{Learning algorithm}
	\label{sec:algoapp}
	Inspired by machine learning practice, we use a modified version of the Stochastic Gradient Descent (SGD) algorithm, capable of dealing with strictly positive weights, to train our single neuron models (see Algo.~\ref{alg:SGD}). 
	To constrain the synapses to be positive during the learning dynamics, at each gradient step, we reset negative synaptic weights to zero. The SGD algorithm minimizes a differentiable objective function, and a common choice in machine learning is the cross-entropy (CE) loss. For binary outputs, the CE loss is given by:
	$
	\mathcal{L}_{\text{CE}}(\boldsymbol{W};\gamma_{ce},\theta_d,\theta_s) = \sum_{\mu=1}^P f_{\gamma_{ce}}\left( \sigma^{\mu}\Delta^{\mu}(\boldsymbol{W};\theta_d,\theta_s) \right)
	$
	where $f_{\gamma_{ce}}(x)=-\frac{x}{2} + \frac{1}{2\gamma_{ce}}\log\left(2\cosh(\gamma_{ce} x)\right)$ and the output pre-activation is given by $\Delta^{\mu}(\boldsymbol{W};\theta_d,\theta_s)=\frac{1}{\sqrt{N}} \sum_{i=1}^N W_i \xi_i - T \sqrt{N}$ for the linear neuron and \eqref{eq:inputsoma} for the nonlinear neuron. The parameter $\gamma_{ce}$ governs the shape of the CE loss function and, consequently, the training robustness, as discussed in~\cite{baldassi2019shaping}.
	
	Note that unlike in multi-layer networks, SGD in a single neuron model leads to a local learning rule, that could be plausibly implemented in a biological neuron, provided an error signal is available. For a single presented pattern $\mu$, a weight $w_{il}$ is changed by an amount proportional to $\delta w_{il}=-\sigma^{\mu}f'_{\gamma_{ce}}(\sigma^{\mu}\Delta^{\mu})g'(\lambda_l^{\mu})\xi_{il}^{\mu}$. This can be interpreted as a `three factor rule', where $-\sigma^{\mu}f'_{\gamma_{ce}}$ is a `soft' error signal available to the whole neuron (possibly a `plateau potential' triggered by apical inputs), $g'(\lambda_l^{\mu})$ is a local, NMDA mediated, dendritic signal, and $\xi_{il}^{\mu}$ is the presynaptic activity.
	\begin{algorithm}[H]
		\caption{SGD with CE loss and positive weights}
		\label{alg:SGD}
		\begin{algorithmic}
			\State \textbf{Hyperparameters:} learning rate $\zeta$, cross-entropy parameter $\gamma_{ce}$
			\For{$t=1,2,\dots$}
			\State $\xi^{\mu},\sigma^{\mu}\leftarrow\text{sample pattern}$ 
			\State 
			$\delta w_{il} \leftarrow \nabla_{w_{il}} \mathcal{L_{\text{CE}}}\left(w_{il}; \xi^{\mu}, \gamma_{ce}\right)$
			\State 
			$w_{il}\leftarrow w_{il} - \zeta \cdot \delta w_{il}$
			\If{$w_{il}<0$}
			\State 
			$w_{il}\leftarrow0$
			\EndIf 
			\EndFor 
		\end{algorithmic}
	\end{algorithm}

	\subsection{Numerical Experiments}
	
	We provide details and algorithmic considerations used for the numerical experiments reported in the paper for both the linear and non-linear neuron models.
	
	For the non-linear neuron model, the hyper-parameters are the two thresholds $\theta_d$ and $\theta_s$, the learning rate $\zeta$, and the CE robustness parameter $\gamma_{ce}$ (for the SGD algorithm). For the linear neuron model, there are only two parameters: the learning rate $\zeta$ and the robustness parameter $\gamma_{ce}$ (for SGD).
	In the linear case, the threshold $\theta_d$ governs the synaptic mean value but does not alter the dynamics with a suitable rescaling of the learning rate with $\theta_d$.
	In the algorithmic capacity simulations, the learning rate is adjusted adaptively: whenever the error stops decreasing, the learning rate is reduced by a factor of two. This process is repeated until the learning rate reaches \( \frac{1}{4096 \cdot N} \), which is on the order of \( 10^{-6} \).
	Otherwise, we perform a simple exponential annealing of the learning rate: at epoch $t$, the learning rate is $\zeta{t}=\zeta(1-d\zeta)^{t}$ with $d\zeta=10^{-4}$. 
	Learning rate decay is justified by the fact that in the linear neuron case, adapting~\cite{rosenblatt1962, minsky1988} convergence proof, one can demonstrate that the perceptron algorithm converges below the critical capacity, provided that the variation in weights at each step is smaller than a certain critical value $dw_c > 0$~\cite{Brunel2004}.
	During single-neuron training, we present one input pattern at a time (the \emph{minibatch size} is $1$). We randomly shuffle the pattern sequence at each dataset presentation (\emph{epoch}).
	Input and output coding levels are fixed to $f_{\text{in}}=f_{\text{out}}=0.5$.
	
	
	Synapses are initialized uniformly at random between zero and twice the theoretical expected value of the mean weight $\bar{w}=\frac{\theta}{f_{\text{in}}N}$. This is the expected value for both the linear and non-linear neurons with a generic activation function and a generic value of the $\theta_s$ threshold in the symmetric $f_{\text{in}}=f_{\text{out}}=0.5$ case, as analytically shown in~Appendix~\ref{sec::Anal}.
	If synapses turn negative during training, a hard boundary condition is enforced, and they are immediately reset to zero.
	For the non-linear neuron, the SGD update rule (Algo.\ref{alg:SGD}) with the cross-entropy loss
	is performed only on the dendritic branches of the first synaptic layer.
	
	For the linear neuron trained with the SGD algorithm, the dynamics is invariant with respect to the rescaling by a positive constant $c$ of the three hyper-parameters: $\zeta$, $\gamma_{ce}$, and $\theta_d$, i.e.: $\zeta\leftarrow \zeta/c$, $\gamma_{ce}\leftarrow c\gamma_{ce}$, $\theta_d\leftarrow \theta_d/c$.
	As a result, the dynamics effectively depends only on two of these hyper-parameters. 
	
	\subsubsection{Hyper-parameter selection}
	
	We report the hyper-parameter selection used for numerical simulations of the non-linear neuron (a.k.a. tree committee machine) and the linear one (a.k.a. perceptron).
	In the storage case, the input size is $N=999$, and the number of dendritic branches of the non-linear neuron is $K=27$.
	For numerical simulations on real-world datasets, the input size is twice the number of pixels in each image, i.e., $N=1568$ for MNIST and FashionMNIST, and $N=6144$ for CIFAR10. The number of dendritic branches is $K=7$ for MNIST and FashionMNIST, and $K=8$ for CIFAR10.
	The Polsky dendritic nonlinear transfer function has biologically estimated parameters $x_{\text{min}}=0.33$ and $\gamma=15$. The non-linear neuron behavior regarding relevant observables is studied as a function of the dendritic and somatic thresholds $\theta_d$ and $\theta_s$.
	
	To optimize neuronal computational performances on relevant observables, we perform a grid search for SGD on the learning rate $\zeta \in \{0.0001, 0.001, 0.01, 0.1, 1.0\}$ and the cross-entropy robustness parameter $\gamma_{ce} \in \{0.001, 0.01, 0.1, 1.0, 10.0, 100.0\}$. 
	
	
	\subsection{Real datasets definition and binarization}
	
	It is necessary to define a binarization procedure for real-world classification datasets (MNIST, FashionMNIST, CIFAR10) such that different classes have comparable input coding levels. Otherwise, as can be observed even in a simple teacher-student setting, the classification task becomes trivial due to the inherent correlation between neuronal classification and coding level, arising from the strictly positive nature of both input patterns and synapses.
	
	One possible choice is to binarize the patterns of real-world datasets into zeros and ones, with a binarization threshold determined by the median value of each pattern. 
	However, there still is an ambiguity in this approach: we can assign zero to pixels below the median and one to pixels above it, or vice versa. These two choices may potentially result in different input pattern coding levels when patterns are imbalanced.
	To address both this binarization ambiguity and trivial correlations between neuronal classification and coding level, a possible approach is to integrate both binarization types within the same pattern.
	Consequently, both the input size and the number of synaptic weights for each neuron double. Instances of neurons being active in the absence of input have been observed, for example, in~\cite{stimulus-absence}. To further reduce intra-branch correlations, pixels in each image are shuffled using the same random permutation prior to binarization. Moreover, to maintain the input locality to the dendritic branch, each branch receives a different, non-overlapping portion of the pattern that combines both binarization choices.
	In the case of Fashion-MNIST, we filter the patterns that have strictly positive median to avoid extremely unbalanced input coding level cases. However, such cases are unavoidable in the MNIST dataset due to the black-digit-on-white-background nature of the dataset.

	\onecolumngrid 
	\section{Analytical results}\label{sec::Anal}
	
	\subsection{Definition of the non-linear model of the neuron}
	
	We recall here the main definitions of the single neuron model studied in the main text of the paper. Given an activity pattern  $\boldsymbol{\xi}^\mu \in \left\{0, 1\right\}^N$, the output of our model of neuron is obtained in two steps. Firstly, the activity pattern is processed by the corresponding dendritic branch; we suppose here that we have $l = 1\,, \dots \,, N/K$ dendritic branches each having a set of $i = 1\,, \dots \,, N$ positive synaptic weights $W_{li}$. The output activity $\tau_l^\mu$ of a given branch $l$ that corresponds to the activity pattern $\boldsymbol{\xi}^\mu$ is obtained as
	\begin{equation}
		\tau_l^\mu =  g\left(\sqrt{\frac{K}{N}} \sum_{i=1}^{N/K} W_{li} \xi_{li}^\mu - \sqrt{\frac{N}{K}}\theta_{d}\right) \equiv  g\left(\lambda^\mu\right)
	\end{equation}
	where $\theta_{d}$ is a threshold modeling inhibition at the level of the dendritic branch, while $g(\cdot)$ is a generic positive, (possibly) non-linear function.
	Secondly, the output of each branch is combined linearly by using another set of $K$ synaptic weights $c_l$, $l = 1\,, \dots \,, K$ and the output is obtained as
	\begin{equation}
		\label{eq:neuron_supp}
		\sigma^{\mu}_{\text{out}}
		= \Theta\left[\frac{1}{\sqrt{K}}\sum_{l=1}^{K} c_{l} \, \tau^{\mu}_{l} - \sqrt{K} \theta_s \right]
	\end{equation}
	where $\Theta(x)$ is the Heaviside theta function that is $1$ if $x>0$ and 0 otherwise. The parameter $\theta_s$ is a threshold modelling inhibition coming from inhibitory neurons. In the following we will consider, for simplicity $c_l = 1$ for every $l = 1\,, \dots\,, K$. 
	
	\subsection{Training set and partition function}
	We consider a training set composed of $P = \alpha N$ random i.i.d. activity patterns $\boldsymbol{\xi}^{\mu}\in \{0,1\}^N$ and i.i.d. labels $\sigma^{\mu} \in \lbrace 0, 1\rbrace$ with $\mu = 1\,, \dots \,, P$. The probability distribution of each component of a pattern is given by
	\begin{equation}
		\label{eq::P(xi)}
		P(\xi_{li}^\mu) = f_{\text{in}} \, \delta\!\left(\xi_{li}^\mu - 1\right) + (1- f_\text{in}) \, \delta\!\left(\xi_{li}^{\mu}\right)
	\end{equation}
	where $f_{\text{in}}$ is the \emph{input coding level} of the patterns. We consider a probability distribution of labels to be equal in form to~\eqref{eq::P(xi)} but we allow the possibility to have different coding level in the output $f_{\text{out}}$. 
	
	In order to study the volume of synaptic weights that correctly associate to a given pattern of activity $\boldsymbol{\xi}^\mu$ the corresponding label $\sigma^\mu$ we use a standard statistical mechanics approach~\cite{gardner1988The,gardner1988optimal}. Firstly, we define 
	the characteristic function 
	\begin{equation}
		\label{eq:X}
		X_{\xi,\sigma} (W) = \prod_{\mu} \Theta \left(\frac{\left( 2\sigma^{\mu}-1\right)}{\sqrt{K}}\left(\sum_{l=1}^{K} c_{l} \, \tau^{\mu}_{l} - K \theta_s \right)-\kappa\right)	
	\end{equation}
	which is $1$ when a given weight $W_{li}$ correctly classifies all the patterns (we will call this a \emph{solution}), and $0$ otherwise. The volume of the allowed synapses, which in statistical mechanics is known as the \emph{partition function}, is therefore:
	\begin{equation}
		\label{eq:Z}
		Z = \int d\mu(W) \, X_{\xi,\sigma}(W)
	\end{equation}
	where $d\mu (W)$ is the measure over the weights. We will consider in the following
	\begin{equation}
		\int d \mu(W) \, \bullet \equiv  \int_0^\infty \prod_{li} d W_{li} \, \bullet
	\end{equation}
	without giving constraints to the norm of the weights. As we will see the norm will be imposed self-consistently by the learning problem.
	
	\subsection{Replica method}\label{sec::replica_method}
	To compute the average entropy $\langle \ln Z \rangle_{\xi, \sigma }$ of synaptic weights solutions in the large $N$ limit, we resort the Replica Method \cite{mezard1987spin} that is based on the following identity:
	\begin{equation*}
		\langle \ln Z \rangle_{\xi, \sigma } = \lim_{n \to 0} \frac{ \langle  Z^{n} \rangle_{ \xi, \sigma} - 1}{n} = \lim_{n \to 0} \frac{1}{n}\ln \langle  Z^{n} \rangle_{ \xi, \sigma}
	\end{equation*}
	This trick reconducts the problem of estimating the log of the partition function in~\eqref{eq:Z} to the computation of the average of $n$ independent copies of the systems with the same realization of the disorder of the activity patterns and labels $\boldsymbol{\xi}^{\mu}, \sigma^{\mu}$:
	\begin{equation}
		\langle  Z^{n} \rangle = \left\langle \int \prod_{a=1}^{n}  d\mu (W^{a}) \, \prod_{a,\mu} \theta \left( \frac{\sigma^{\mu}}{\sqrt{K}} \left(\sum_{l=1}^{K} c_l \, g  \left(\sqrt{\frac{K}{N}} \sum_{i=1}^{N/K} W^{a}_{li} \xi^{\mu}_{li} - \sqrt{\frac{N}{K}} \theta_d\right) - K \theta_s \right)-\kappa \right) \right\rangle_{ \xi, \sigma} \,.
	\end{equation}
	We will denote from now on $a$ and $b$ as the index that run over replicas $a\,, b = 1, \dots\,, n$. Notice also that, because of~\eqref{eq:X} we can safely consider having labels $\sigma^\mu = \pm 1$ with the same output coding level as before. The computation follows standard steps \cite{gardner1988The, engel-vandenbroek}, which we will sketch here. Firstly we need to perform the average over the activity patterns $\boldsymbol{\xi}^\mu$; we can do that by introducing the auxiliary variables 
	\begin{equation}
		\lambda_l^{\mu a} = \sqrt{\frac{K}{N}} \sum_{i=1}^{N/K} W_{li} \xi_{li}^\mu - \sqrt{\frac{N}{K}}\theta_{d}
	\end{equation}
	and the corresponding conjugated variables $\hat\lambda_l^{\mu a}$ that arise when we insert the integral representation of the Dirac delta function. The replicated partition function is
	\begin{equation}
		\label{eq:Zn}
		\begin{split}
			\langle  Z^{n} \rangle &= \mathbb{E}_{\sigma} \int \prod_{a}  d\mu (W^{a}) \, \int \prod_{\mu a l} \frac{d \lambda^{\mu a}_l d \hat\lambda^{\mu a}_l}{2\pi} e^{i \lambda_l^{\mu a} \hat \lambda_l^{\mu a}} \prod_{a,\mu} \theta \left( \frac{\sigma^{\mu}}{\sqrt{K}} \left(\sum_{l=1}^{K} c_l \, g  \left(\lambda_l^{\mu a}\right) - K \theta_s \right) -\kappa\right) e^{i \sqrt{\frac{N}{K}} \theta_d \sum_{\mu a l} \lambda_{l}^{\mu a}}. \\
			& \times \prod_{li \mu} \left\langle e^{- i \xi_{li}^\mu \sqrt{\frac{K}{N}} \sum_a W_{li}^a \hat \lambda_{l}^{\mu a}} \right\rangle_{\xi_{li}^\mu}\,.
		\end{split}
	\end{equation} 
	The average over patterns can now be performed. In the large $N$ limit, we can use the central limit theorem, having
	\begin{equation}
		\begin{split}
			\prod_{li \mu} \left\langle e^{- i \xi_{li}^\mu \sqrt{\frac{K}{N}} \sum_a W_{li}^a \hat \lambda_{l}^{\mu a}} \right\rangle_{\xi_{li}^\mu} &= \prod_{li \mu}\left[ 1 - f_{\text{in}} + f_{\text{in}} e^{- i \sqrt{\frac{K}{N}} \sum_a W_{li}^a \hat \lambda_{l}^{\mu a}} \right] \simeq \\
			&= \prod_{li \mu} e^{- i f_{\text{in}} \sqrt{\frac{K}{N}} \sum_a W_{li}^a \hat \lambda_{l}^{\mu a} - \frac{f_{\text{in}}(1-f_{\text{in}}) K}{2N} \left(\sum_a W_{li}^a \hat{\lambda}_l^{\mu a}\right)^2} \\
			&= e^{- i f_{\text{in}} \sqrt{\frac{K}{N}} \sum_{\mu a l}\hat \lambda_{l}^{\mu a} \sum_i W_{li}^a - \frac{f_{\text{in}}(1-f_{\text{in}}) K}{N} \sum_{\mu l} \sum_{a<b} \left( \sum_i W_{li}^a W_{li}^b \right) \hat \lambda_{l}^{\mu a} \hat \lambda_{l}^{\mu b} }\\
			&\times e^{- \frac{f_{\text{in}}(1-f_{\text{in}}) K}{2N} \sum_{\mu a l}  \sum_i \left( W_{li}^a\, \hat \lambda_{l}^{\mu a}\right)^2 }	\end{split}
	\end{equation}
	By defining appropriate order parameters, it is possible to conveniently study the problem in the large-$N$ limit. We define:
	\begin{subequations}\label{eq:ordpar_comm}
		\begin{align}
			&\sum\limits_{i} W^{a}_{li} = \frac{N}{K}\overline{W}+\sqrt{\frac{N}{K}}M^{a}_{l}\label{eq:M}\\
			&q^{ab}_{l} \equiv \frac{K}{N}\sum\limits_{i} W^{a}_{li}W^{b}_{li} \label{eq:qab}\\
			&Q^{a}_{l} \equiv \frac{K}{N} \sum\limits_{i}\left(W^{a}_{li}\right)^{2} \label{eq:Qab}
		\end{align}
	\end{subequations}
	\eqref{eq:M} represents the average synaptic weight; we expressed it in two contributions. The first one represents an averaged \emph{scaled} synaptic weight
	\begin{equation}
		\overline{W} = \frac{\theta_d}{f_{\text{in}}}
	\end{equation}
	justified by the fact that for each sub-perceptron of the first layer only $\frac{N}{K} f_{\text{in}}$ synapses contribute. The second term instead is a $\sqrt{\frac{N}{K}}$ correction that is needed in order to fine tune the average synaptic weight relatively to the threshold $\theta_d$. 
	
	The quantity $q^{ab}_l$ is the overlap between the weights of two different replicas $a$ and $b$ belonging to the same dendritic branch $l$, while 
	$Q^{a}_l$ represents the averaged squared norm of a synaptic weight belonging to dendritic branch $l$.

	Enforcing the definitions~\eqref{eq:ordpar_comm} in~\eqref{eq:Zn}, by using Dirac delta functions, we can express the replicated partition function as an integration over the order parameters $q_l^{ab}, \hat{q}, Q, \hat{Q}, M, \hat{M}$:
	\begin{equation}
		\label{eq:ZnSP}
		\begin{aligned}
			\langle Z^{n} \rangle_{\xi, \sigma} =& \int\prod\limits_{a<b,l} \frac{dq^{ab}_{l} d\hat{q}^{ab}_{l}}{2\pi K/N} \int\prod\limits_{a,l} \frac{dQ^{a}_{l}d\hat{Q}^{a}_{l}}{2\pi K/N}\int\prod\limits_{a,l} \frac{dM^{a}_{l}d\hat{M}^{a}_{l}}{2\pi\sqrt{K/N}} \, e^{\,-\frac{N}{K}\sum\limits_{a<b,l}q^{ab}_{l}\hat{q}^{ab}_{l}\,-\,\frac{N}{K}\sum\limits_{a,l}Q^{a}_{l}\hat{Q}^{a}_{l}\, -\frac{N}{K}\overline{W}\sum\limits_{a,l}\hat{M}^{a}_{l}}\\
			&\times e^{\frac{N}{K}\,G_{S} \left(\hat{q}^{ab}_{l},\hat{Q}^{a}_{l},\hat{M}^{a}_{l}\right) \,+\, N\,\alpha\, G_{E} \left(q^{ab}_{l}, Q^{a}_{l}, M^{a}_{l} \right)}.
		\end{aligned}
	\end{equation}
	where we collected the entropic contribution $G_S$ and the energetic one $G_E$. The first is the usual term that counts how many coupling vectors $W^a$ fulfill the constraints \eqref{eq:ordpar_comm}; the second is specific to the learning rule which is used, and depends on the Heaviside function that counts learned patterns:
	\begin{subequations}
		\begin{align}\label{eq:Gs}
			G_{S}\left(\hat{q}^{ab}_{l},\hat{Q}^{a}_{l},\hat{M}^{a}_{l}\right) &= \ln \int\limits_{0}^{\infty}\prod\limits_{a,l} dW^{a}_{l} \, e^{\, \sum\limits_{a<b,l} \hat{q}^{ab}_{l}W^{a}_{l} W^{b}_{l} \,+\,\sum\limits_{a,l} \hat{Q}^{a}_{l} (W^{a}_{l})^{2} \,+\,\sum\limits_{a,l}\hat{M}^{a}_{l}W^{a}_{l}}\\
			\label{eq:Ge}
			G_{E}\left(q^{ab}_{l},Q^{a}_{l},M^{a}_{l}\right) &= \ln \E_\sigma\int\prod\limits_{a,l}\frac{d\lambda^{a}_{l}d\hat{\lambda}^{a}_{l}}{2\pi} \Theta\left(\frac{\sigma}{\sqrt{K}} \left( \sum_l c_l \, g(\lambda_{l}^a) - K \theta_s\right)-\kappa \right) \, e^{i \sum\limits_{a,l} \lambda^{a}_{l}\hat{\lambda}^{a}_{l} -f_{\text{in}}(1-f_{\text{in}})\sum\limits_{a<b,l} q^{ab}_{l}\hat{\lambda}^{a}_{l}\hat{\lambda}^{b}_{l}}\\
			&\times e^{-\frac{f_{\text{in}}(1-f_{\text{in}})}{2}\sum\limits_{a,l}Q^{a}_{l}(\hat{\lambda}^{a}_{l})^{2}\,-\,if_{\text{in}}\sum\limits_{a,l}M^{a}_{l}\hat{\lambda}^{a}_{l}}. \nonumber
		\end{align}
	\end{subequations}
	We can now evaluate in the large $N$ limit~\eqref{eq:ZnSP} using the saddle point method. In order restrict the space where to search saddle points we proceed by assuming a particular form of the order parameters, which is the main topic of the next section.
	
	\subsection{Replica Symmetric analysis}
	We use a Replica Symmetric (RS) ansatz, i.e. we assume that the order parameters do not depend on the replica indexes and of the index corresponding to the dendritic branch:
	\begin{subequations}
		\begin{align}
			&q_l^{ab} = q\,, \qquad Q_l^a = Q\,, \qquad M_l^a = M \,,\\
			&\hat q_l^{ab} = \hat q\,, \qquad \hat Q_l^a = \hat Q\,, \qquad \hat M_l^a = \hat M
		\end{align}
	\end{subequations}
	In the RS ansatz and in the small $n$ limit, using \textit{Hubbard-Stratonovich} transformations:
	\begin{equation*}
		e^{\frac{1}{2}b x^2} = \int \frac{dz}{\sqrt{2\pi}} \, e^{-\frac{z^2}{2} + \sqrt{b}x z}\,,
	\end{equation*}
	the entropic and energetic terms are the following
	\begin{subequations}
		\begin{align}
			& \mathcal{G}_S \equiv \lim\limits_{n\to0}\frac{1}{nK} G_S(\hat{q}, \hat{Q}, \hat{M}) = \ln \sqrt{\frac{2\pi}{\hat{q}-2\hat{Q}}} + \frac{1}{2}\left(\frac{\hat{M}^{2}+\hat{q}}{\hat{q}-2\hat{Q}}\right)+\int Dz \, \ln H \left(-\frac{\hat{M}+\sqrt{\hat{q}}z}{\sqrt{\hat{q}-2\hat{Q}}}\right)\label{Gs}\\
			&   \mathcal{G}_E \equiv \lim\limits_{n\to0}\frac{G_{E}\left( q,Q,M \right)}{n} = \, \underset{\sigma}{\E} \int\prod_{l} Dt_{l} \, \ln \left[
			\int \prod_{l} D\lambda_{l} \, \Theta\left(\frac{\sigma}{\sqrt{K}}\left(\sum\limits_{l} c_{l} \, g\left(\sqrt{f_{\text{in}}(1-f_{\text{in}})(Q-q)}\lambda_{l} + a_{l}\right)-K\theta_{s}\right)-\kappa\right)\right] \label{Ge}
		\end{align}
	\end{subequations}
	where we have introduced the variable
	\begin{equation}
		a_l \equiv f_\text{in} M+\sqrt{f_\text{in}(1-f_\text{in})q}\,t_{l}
	\end{equation}
	for convenience. In~\eqref{Gs} we have also introduced $H(x) \equiv \int_{x}^{\infty} Dz = \frac{1}{2} \text{Erfc}\left(\frac{x}{\sqrt{2}}\right)$ and $Dz \equiv G(z) \, dz$ with $G(z)$ being a standard normal Gaussian $G(z) = \exp (-z^2/2)/\sqrt{2\pi}$.
	
	\subsubsection{Large $K$ limit}
	
	We focus on the limit $K \to \infty$ $\left(\text{with}\,\,\, \frac{K}{N} \to 0\right)$, for two main reasons: 
	\begin{itemize}
		\item on the analytical level, it allows to simplify the numerical evaluation of the saddle point equations corresponding to the RS ansatz;
		\item it is \emph{biologically} realistic: the number of dendritic branches in neurons is typically large (in some cases even more than a hundred, REF) and the number of synapses in each branch is typically large as well (REFS?).
	\end{itemize}
	To evaluate this limit, we need to do some manipulations on the energetic contribution \eqref{Ge} which are based on the central limit theorem. Let us first consider the term in square brackets:
	\begin{equation}\label{form:I}
		\begin{aligned}
			I &= \int \prod_{l} D\lambda_{l}\, \Theta \left(\frac{\sigma}{\sqrt{K}}\left(\sum\limits_{l} c_{l} \, g \left(\sqrt{f_{\text{in}}(1-f_{\text{in}})(Q-q)} \lambda_{l}+a_{l}\right)-K\theta_{s}\right)-\kappa\right)\\
			& = \int \frac{dh \, d\hat{h}}{2\pi} e^{-ih\hat{h}} \, \Theta \left[\sigma \left(h-\sqrt{K}\theta_{s}\right)-\kappa\right] \int \prod_{l} D\lambda_{l} \, e^{\frac{i\hat{h}}{\sqrt{K}}\sum\limits_{l} c_{l} \, g\left(a_{l}+ \sqrt{f_{\text{in}}(1-f_{\text{in}})(Q-q)}\, \lambda_{l}\right)}
		\end{aligned}
	\end{equation}
	In the large $K$ limit we can therefore expand the exponential up to second order
	\begin{equation}
		\begin{aligned}
			I &\simeq \int \frac{dh \, d\hat{h}}{2\pi} e^{-ih\hat{h}} \, \Theta \left(\sigma (h-\sqrt{K}\theta_{s})-\kappa\right) \int \prod_{l} D\lambda_{l} \, \left[1+\frac{i\hat{h}}{\sqrt{K}}\sum\limits_{l} c_{l} \, g\left(a_{l}+\sqrt{f_{\text{in}}(1-f_{\text{in}})(Q-q)}\, \lambda_{l}\right)+\right.\\
			& \left.\qquad\qquad\qquad\qquad\qquad\qquad\qquad\qquad\qquad\qquad\qquad\qquad\qquad\qquad-\frac{\hat{h}^{2}}{2K}\left(\sum\limits_{l} c_{l} \, g\left(a_{l}+\sqrt{f_{\text{in}}(1-f_{\text{in}})(Q-q)}\, \lambda_{l}\right)\right)^{2}\right].
		\end{aligned}
	\end{equation}
	and we can integrate with respect to all the $\lambda_l$ variables term by term. Exponentiating the expression again and integrating over $\hat h$ we get
	\begin{equation}
		I = \int Dh \; \Theta
		\left(\sigma(M^{(0)}\,+\,\sqrt{\Delta^{(0)}}h\,-\,\sqrt{K}\theta_s)-\kappa\right)
	\end{equation}
	where we have introduced the variables:
	\begin{subequations}\label{form:md}
		\begin{align}
			& M^{(0)} = \frac{1}{\sqrt{K}}\sum\limits_{l} c_{l} \langle g \rangle_{\lambda}\\
			& D^{(0)} = \frac{1}{K}\sum\limits_{l} c^{2}_{l} \left[\,\langle g^{2} \rangle_{\lambda} - \langle g \rangle^{2}_{\lambda}\,\right].
		\end{align}
	\end{subequations}
	and the notation 
	\begin{equation}
		\langle g \rangle_{\lambda} = \int D \lambda \, g\left(f_\text{in} M+\sqrt{f_\text{in}(1-f_\text{in})q}\,t_{l}+\sqrt{f_{\text{in}}(1-f_{\text{in}})(Q-q)}\, \lambda \right) \,.
	\end{equation}
	Using again the central limit theorem for the $K$ integrals over the variable $t_{l}$ we get:
	\begin{equation}\label{form:eneg}
		\mathcal{G}_{E}= \underset{\sigma}{\E} \int Dt \ln \int D\lambda \, \Theta \left[\sigma\left(M_0\,+\, \sqrt{D_{0}}\,t\,+\sqrt{D_{1}}\,\lambda\,-\,\sqrt{K}\theta_{s}\right)-\kappa\right]
	\end{equation}
	where:
	\begin{subequations}
		\begin{align}
			& M_0 = \frac{1}{\sqrt{K}}\, \sum\limits_{l} c_{l} \langle\, \langle g \rangle_{\lambda} \rangle_{t} = m_{c} \langle\, \langle g \rangle_{\lambda} \rangle_{t}\label{Mord}\\ 
			& D_{0} = \frac{1}{K}\sum\limits_{l} c^{2}_{l} \left[\langle \, \langle g \rangle^{2}_{\lambda}\rangle_{t} \, -\, \langle \, \langle g \rangle_{\lambda} \rangle^{2}_{t} \right] = w_{c} \left[\langle \, \langle g \rangle^{2}_{\lambda}\,\rangle_{t} \, -\, \langle \,\, \langle g \rangle_{\lambda} \rangle^{2}_{t} \right]\\
			& D_{1} = w_{c} \left[\langle \, \langle g^{2} \rangle_{\lambda}\rangle_{t} \,-\, \langle \, \langle g \rangle^{2}_{\lambda}\rangle_{t}\right]
		\end{align}
	\end{subequations}
	and for example
	\begin{equation}
		\langle \langle g \rangle_{\lambda}\rangle_{t} = \int D t D \lambda   \, g\left(f_\text{in} M+\sqrt{f_\text{in}(1-f_\text{in})q} \, t +\sqrt{f_{\text{in}}(1-f_{\text{in}})(Q-q)}\, \lambda \right) \,.
	\end{equation}
	Using the definition of the committee machine in which all weights in the second layer are s.t.  $c_l=1$, we have $m_c = \sqrt{K}$ and $w_c = 1$.  In order to have a well-defined large $K$ limit we have to impose (analogously to what we have done on the dendritic threshold $\theta_d$) that the divergence induced by the somatic threshold $\theta_s$ cancels with the one coming from $M_0$. We therefore impose that $M$ scales, in the large $K$ limit, as
	\begin{equation}\label{eq:scalingM}
		M = \overline{M} + \frac{\delta M}{\sqrt{K}} \,.
	\end{equation}
	$M_0$ can be simplified by making a rotation over the integration measures $\lambda$ and $t$
	\begin{equation}
		\begin{split}
			M_0 = \sqrt{K}\langle \, \langle g \rangle_{\lambda} \rangle_{t}& = \sqrt{K} \int D\lambda \, Dt \, g \left( \sqrt{f_{\text{in}}(1-f_{\text{in}})q}\,t + \sqrt{f_{\text{in}}(1-f_{\text{in}})(Q-q)} \lambda + fM \right)\\
			&=\sqrt{K}\int D\lambda \, g\left(\sqrt{f_{\text{in}}(1-f_{\text{in}})Q}\, \lambda + f_{\text{in}}M \right)
		\end{split}
	\end{equation}
	We can now insert the scaling in~\eqref{eq:scalingM}
	\begin{equation}
		\begin{aligned}
			M_0 &= \sqrt{K}\int D\lambda\, g\left(\sqrt{f_{\text{in}}(1-f_{\text{in}})Q}\, \lambda + f_{\text{in}}M \right)\\
			& \simeq \sqrt{K}\int D\lambda \, g\left( \sqrt{f_{\text{in}}(1-f_{\text{in}})Q}\lambda + f_{\text{in}}\overline{M}\right)\,+f_{\text{in}} \, \delta M \int D\lambda \, g'\left(\sqrt{f_{\text{in}}(1-f_{\text{in}})Q}\, \lambda + f_{\text{in}}\overline{M}\right)\\
			&=\sqrt{K}\,\theta_{s} + \Delta
		\end{aligned}
	\end{equation}
	Therefore, $\overline{M}$ is fixed by the relation:
	\begin{equation}\label{eq:equation_for_Mb}
		\theta_{s} = \int Dy \, g\left( \sqrt{f_{\text{in}}(1-f_{\text{in}})Q}\,y + f_{\text{in}}\overline{M}\right) \,,
	\end{equation}
	that involves the output threshold. Hence the energetic term \eqref{form:eneg} becomes:
	\begin{equation}\label{eq:GEfinal}
		\mathcal{G}_{E} = \underset{\sigma}{\E} \int Dz \, \ln H \left(\frac{\kappa - \sigma\Delta +\sqrt{D_{0}}\,z}{\sqrt{D_{1}}}\right) = \int Dz \, \left[ f_{\text{out}} \ln H\left(\frac{\kappa- \Delta +\sqrt{D_{0}}\,z }{\sqrt{D_{1}}}\right) + (1-f_{\text{out}}) \ln H\left(\frac{\kappa + \Delta +\sqrt{D_{0}}\,z}{\sqrt{D_{1}}}\right) \right]
	\end{equation}
	where the order parameters, strictly dependent on the choice of the activation function $g(x)$, are simplified as:
	\begin{subequations}
		\begin{align*}
			\Delta & \equiv f_{\text{in}} M \int Dx \, g'\left( \sqrt{f_{\text{in}}(1-f_{\text{in}}) Q} \, x + f_{\text{in}} \overline{M} \right) \\
			D_0 & = \Delta_q - \Delta_0\\
			D_1 &= \Delta_Q - \Delta_q
		\end{align*}
	\end{subequations}
	where we have renamed $\delta M$ by $M$ with a slight abuse of notation. We have also defined the generic ``effective order parameter'' or \emph{kernel} functions
	\begin{equation}
		\label{eq::effective_q}
		\Delta_q = \int Dx \left[ \int D y \, g\left( \sqrt{f_{\text{in}}(1-f_{\text{in}}) q} \, x + \sqrt{f_{\text{in}}(1-f_{\text{in}}) (Q-q)} \, y +  f_{\text{in}} \overline M   \right)\right]^2
	\end{equation}
	$\Delta_Q$ and $\Delta_0$ being obtained by simply substituting in the previous expression $q \to Q$ and $q \to 0$ respectively. We report them here for clarity
	\begin{subequations}
		\begin{align}
			\Delta_Q &= \int Dx \, g^2\left(\sqrt{f_{\text{in}}(1-f_{\text{in}}) Q} \, x + f_{\text{in}} \overline M  \right) \,,\\
			\Delta_0 &= \left[ \int D y \, g\left(\sqrt{f_{\text{in}}(1-f_{\text{in}}) Q} \, y + f_{\text{in}} \overline M   \right)\right]^2\,.
		\end{align}
	\end{subequations}
	We have called $\Delta_q$ an \emph{effective order parameter} since~\eqref{eq:GEfinal} (and therefore the whole quenched entropy) is perfectly equivalent to the one found in the perceptron model studied by Brunel in a series of papers~\cite{Brunel2004,Brunel_2016}, but where order parameters $q$ and $Q$ are substituted respectively by $\Delta_q - \Delta_0$ and $\Delta_Q - \Delta_0$. 
	
	\subsubsection{Free entropy and saddle point equations}
	The average \emph{free entropy} of the dendritic model of a neuron is therefore
	\begin{equation}\label{eq:freeentropy}
		\phi =  \lim\limits_{N\to \infty} \frac{1}{N} \langle \ln Z \rangle_{\xi, \sigma} =\frac{q\hat{q}}{2} - Q \hat{Q} - \overline{W}\hat{M} + \mathcal{G}_S + \alpha \,\mathcal{G}_E
	\end{equation}
	We now need to compute the saddle point equations by differentiating \eqref{eq:freeentropy} with respect to the order parameters $Q, q, M, \hat{Q}, \hat{q}, \hat{M}$. The saddle point equations involving the entropic term (i.e. taking derivatives with respect to $\hat M$, $\hat Q$ and $\hat q$) are the same as in the case of the perceptron
	\begin{subequations}
		\label{eq:sp_0}
		\begin{align}
			\overline{W} &= \int Dz \, \frac{\int_{0}^{\infty} dW \, W e^{-(\hat q - 2\hat Q) \frac{W^2}{2} + (\hat M + \sqrt{\hat q} z)W }}{\int_{0}^{\infty} dW \, e^{-(\hat q - 2\hat Q) \frac{W^2}{2} + (\hat M + \sqrt{\hat q} z)W }} \label{eq:sp1}\\
			Q &= \int Dz \, \frac{\int_{0}^{\infty} dW \, W^2 e^{-(\hat q - 2\hat Q) \frac{W^2}{2} + (\hat M + \sqrt{\hat q} z)W }}{\int_{0}^{\infty} dW \, e^{-(\hat q - 2\hat Q) \frac{W^2}{2} + (\hat M + \sqrt{\hat q} z)W }} \label{eq:sp2}\\
			q &= \int Dz \,  \frac{\int_{0}^{\infty} dW \, \left( W^2 - \frac{z W}{\sqrt{q}}\right) e^{-(\hat q - 2\hat Q) \frac{W^2}{2} + (\hat M + \sqrt{\hat q} z)W }}{\int_{0}^{\infty} dW \, e^{-(\hat q - 2\hat Q) \frac{W^2}{2} + (\hat M + \sqrt{\hat q} z)W }} =  \int Dz \, \left[ \frac{\int_{0}^{\infty} dW \, W e^{-(\hat q - 2\hat Q) \frac{W^2}{2} + (\hat M + \sqrt{\hat q} z)W }}{\int_{0}^{\infty} dW \, e^{-(\hat q - 2\hat Q) \frac{W^2}{2} + (\hat M + \sqrt{\hat q} z)W }} \right]^2 \label{eq:sp3}
		\end{align}
	\end{subequations}
	In order to express the remaining saddle point equations in a compact way, we define the quantities:
	\begin{subequations}
		\label{asigma}
		\begin{align*}
			& a_\sigma(z) = \frac{\sqrt{D_{0}} z -\sigma\Delta + \kappa}{\sqrt{D_{1}}} = \sqrt{\frac{D_0}{D_{1}}} (z - \tau_\sigma) \\
			& \tau_\sigma = \frac{\sigma \Delta -\kappa}{\sqrt{D_0}} \,.
		\end{align*}
	\end{subequations}
	Deriving~\eqref{eq:freeentropy} with respect to $M$, $Q$ and $q$ lead respectively to
	\begin{subequations}
		\label{eq:sp_1}
		\begin{align}
			0 &= \mathbb{E}_\sigma \sigma \int Dz \, \frac{G(a_\sigma (z))}{H(a_\sigma(z))} \label{eq:sp4}\\ 
			\hat Q &= \frac{\alpha}{2} \mathbb{E}_\sigma \int Dz \, \frac{G(a_\sigma (z))}{H(a_\sigma(z))} \left[ \frac{a_\sigma(z)}{D_1} \frac{d D_1}{dQ} - \frac{z}{\sqrt{D_0 D_1}} \frac{d D_{0}}{d Q}\right]\label{eq:sp5}\\
			\hat q &= \alpha \mathbb{E}_\sigma \int Dz \, \frac{G(a_\sigma (z))}{H(a_\sigma(z))} \left[ -\frac{a_\sigma(z)}{D_1} \frac{d D_1}{dq} + \frac{z}{\sqrt{D_0 D_1}} \frac{d D_{0}}{d q} \right]\label{eq:sp6}
		\end{align}
	\end{subequations}
	where we used the saddle point equation~\eqref{eq:sp4} when performing the derivative in $Q$\footnote{this is why no derivative with respect to $Q$ of $\Delta$ compares in~\eqref{eq:sp5}}. The six saddle points Eqs.~(\eqref{eq:sp1}, \ref{eq:sp2}, \ref{eq:sp3}, \ref{eq:sp4}, \ref{eq:sp5}, \ref{eq:sp6}), obtained in the large $K,N\to\infty$ limit with $N\gg K$, have to be numerically solved to obtain the values of the order parameters and represent the final result of our RS analysis.
	
	Notice that imposing $g(x) = x$ we recover the previous saddle point expressions obtained for the simple linear neuron model~\cite{Brunel2004,Brunel_2016}.
	Notice also that in the case $f_{\text{out}} = \frac{1}{2}$ saddle point equation~(\ref{eq:sp4}) gives $M=0$; in this case therefore $\tau_\sigma = 0$.

	\subsubsection{Effective order parameters for some non-linearities} 
	We report here the analytical expressions of the effective order parameters for several non-linearities of interest
	\begin{itemize}
		\item \emph{Recovering the one-layer neuron model}: if we impose $g(x) = x$ we recover the one-layer neuron model. We report here the expressions of the corresponding effective order parameters for convenience
		\begin{subequations}
			\label{eq::effective_order_parameters_perceptron}
			\begin{align}
				\Delta &= f_{\text{in}} M \\
				\Delta_q &= f_{\text{in}} (1 - f_{\text{in}}) q + f_{\text{in}}^2 \overline{M}^2
			\end{align}
		\end{subequations}
		and, in particular
		\begin{subequations}
			\begin{align}
				\Delta_Q &= f_{\text{in}} (1 - f_{\text{in}}) Q + f_{\text{in}}^2 \overline{M}^2 \\
				\Delta_0 &= f_{\text{in}}^2 \overline{M}^2
			\end{align}
		\end{subequations}
		As a result $\overline{M}$ does not appear anywhere in the energetic term of equation~\eqref{eq:GEfinal}, and $\theta_s = f_{\text{in}} \overline{M}$ is also irrelevant.   
		\item \emph{Theta non-linearity}: $g(x) = \Theta(x)$.
		
		To evaluate the integrals it is useful to use the following identity
		\begin{equation}
			\int Dz \, H^2\left( a + b z \right) = H\left( \frac{a}{\sqrt{1+b^2}} \right) - 2 T \left( \frac{a}{\sqrt{1+b^2}
			}, \frac{1}{\sqrt{1+2b^2}} \right)
		\end{equation}
		where $T$ is the Owen's $T$ function defined as
		\begin{equation}
			T(h, s) \equiv \frac{1}{2\pi} \int_{0}^{s} dx \, \frac{e^{-(1+x^2)\frac{h^2}{2}}}{1+x^2} \,,
		\end{equation}
		and has the following important properties
		\begin{subequations}
			\begin{align}
				T(h,s) & \simeq \frac{G(h) s}{\sqrt{2\pi}} + O\left(s^2\right) \qquad \text{for} \; s\to 0\\
				T(h,1) & = \frac{1}{2} H(h) H(-h)
			\end{align}
		\end{subequations}
		where we remind that $G(x)$ is the Gaussian with mean zero and unit variance. Defining the quantity
		\begin{equation}
			M_\star = \frac{f_{\text{in}} \overline{M}}{\sqrt{f_{\text{in}}(1-f_{\text{in}}) Q}}
		\end{equation}
		the effective order parameters for the theta non-linearity are
		\begin{subequations}
			\begin{align}
				\Delta & = M_\star G\left( - M_\star \right) \\
				\Delta_q &= H(-M_\star) - 2 T\left( M_\star, \sqrt{\frac{Q-q}{Q+q}} \right)
			\end{align}
		\end{subequations}
		and, in particular
		\begin{subequations}
			\begin{align}
				\Delta_Q &= H(-M_\star)\,, \\
				\Delta_0 &= H(-M_\star)^2 \,.
			\end{align}
		\end{subequations}
		$\overline{M}$ is fixed by the relation
		\begin{equation}
			\theta_s = H(- M_\star) \,.
		\end{equation}
		\item \emph{ReLU non-linearity}: $g(x) = x \Theta(x)$ 
		We have
		\begin{subequations}
			\begin{align}
				\Delta & = f_\text{in} M H\left( - M_\star \right) \\
				\Delta_q &= f_{\text{in}}(1-f_{\text{in}})(Q-q)\sqrt{\frac{Q-q}{Q+q}} G^2\left(\frac{f_{\text{in}} \overline{M}}{\sqrt{f_{\text{in}}(1-f_{\text{in}})(Q+q)}}\right) \\
				&+ f_{\text{in}}\left( (1-f_{\text{in}})q+f_{\text{in}} \overline{M}^2 \right) \left[ H\left(- M_\star\right) -2T\left( - M_\star, \sqrt{\frac{Q-q}{Q+q}} \right) \right] \nonumber \\
				&+2f_{\text{in}}\overline{M}\sqrt{f_{\text{in}}(1-f_{\text{in}})Q} \, G\left( M_\star\right)H\left(- M_\star \sqrt{\frac{Q-q}{Q+q}}\right) + 2f_{\text{in}}(1-f_{\text{in}})q\sqrt{\frac{Q-q}{Q+q}} G\left(M_\star \right)G\left( M_\star \sqrt{\frac{Q-q}{Q+q}}\right) \nonumber
			\end{align}
		\end{subequations}
		and in particular
		\begin{subequations}
			\begin{align}
				\Delta_Q &= f_{\text{in}}\left( (1-f_{\text{in}})Q+f_{\text{in}} \overline{M}^2 \right) H\left(- M_\star\right) + f_{\text{in}}\overline{M}\sqrt{f_{\text{in}}(1-f_{\text{in}})Q} \, G\left(- M_\star\right)\,, \\
				\Delta_0 &= \left[\sqrt{f_{\text{in}} (1 - f_{\text{in}}) Q}\, G\left( M_\star \right)+ f_{\text{in}} \overline{M} H\left(- M_\star\right) \right]^2 
			\end{align}
		\end{subequations}
		Note that $\overline{M}$ is fixed by the relation
		\begin{equation}
			\label{eq:equation_for_Mb_relu}
			\theta_s = \sqrt{f_{\text{in}} (1 - f_{\text{in}}) Q}\, G\left(M_\star \right)+ f_{\text{in}} \overline{M} H\left(- M_\star\right)
		\end{equation}
	\end{itemize}
	
	\subsubsection{Small data regime}
	
	In the $\alpha \to 0$ limit the saddle point equations can be solved exactly. Indeed $\hat q = \hat Q = 0$ and equations~\eqref{eq:sp1}, \eqref{eq:sp2}, \eqref{eq:sp3} give respectively
	\begin{subequations}
		\label{eq:sp_alpha0}
		\begin{align}
			\overline{W} = \frac{\theta_d}{f_{\text{in}}} &= \frac{\int_{0}^{\infty} dW \, W e^{ \hat M W }}{\int_{0}^{\infty} dW \, e^{ \hat M W }} = - \frac{1}{\hat M} \label{eq:sp1_alpha0}\\
			Q &= \frac{\int_{0}^{\infty} dW \, W^2 e^{\hat M W }}{\int_{0}^{\infty} dW \, e^{\hat M W }} = \frac{2}{\hat M^2} \label{eq:sp2_alpha0}\\
			q &= \left[ \frac{\int_{0}^{\infty} dW \, W e^{ \hat M W }}{\int_{0}^{\infty} dW \, e^{ \hat M W }}\right]^2 = \frac{1}{\hat M^2} \label{eq:sp3_alpha0}
		\end{align}
	\end{subequations}

	\subsubsection{Distribution of dendritic preactivations}
	
	We derive here the distribution of dendritic preactivations after learning. 
	Since the each dendritic branch has access to an indepedent portion of the input, the distribution of the preactivations if factorized over the $K$ dendritic branches.
	In the large $K$ limit the distribution tends to a Gaussian $\mathcal{N}(\mu, \sigma)$ as can be inspected from the post activation mean~\eqref{eq:equation_for_Mb} and from the argument of the kernel functions~\eqref{eq::effective_q}. Denoting by $\lambda_l \equiv \sqrt{\frac{K}{N}} \sum_{i=1}^{N/K} W_{li} \xi_{li} - \sqrt{\frac{N}{K}} \theta_d$ as in the main text, the mean and the variance of the distribution of the $l-$th dendritic branch are respectively
	\begin{subequations}
		\begin{align}
			\mu &= \mathbb{E}_{\boldsymbol{\xi}} \lambda_l = f_{\text{in}} \overline{M}\\
			\sigma^2 &= \mathbb{E}_{\boldsymbol{\xi}} \lambda_l^2 - \mu^2 = f_{\text{in}} (1- f_{\text{in}}) Q 
		\end{align}
	\end{subequations}
	as confirmed by~\eqref{eq:equation_for_Mb}.
	
	In the $\alpha \to 0$ the variance above can be expressed explicitly in terms of $\theta_d$ and $f_{\text{in}}$ thanks to~\eqref{eq:sp1_alpha0} and~\eqref{eq:sp2_alpha0}. We get that the standard deviation of the dendritic preactivation depends linearly on the dendritic inhibition threshold
	\begin{equation}
		\sigma = \theta_d \sqrt{ \frac{2(1- f_{\text{in}})}{f_{\text{in}}}} \,.
	\end{equation}
	We checked that this linear relation is satisfied even at finite $\alpha$, or considering a different distribution over the weights at initialization, see below. This shows that if $\theta_d$ is small one does not completely use the non-linearity and the model behaves like a one-layer model.

	\subsubsection{Limit of large somatic thresholds}
	
	When the somatic threshold $\theta_s$ diverges, the only way to satisfy~\eqref{eq:equation_for_Mb} is that $\overline{M}$ diverges if the non-linearity $g$ is unbounded. This can be inspected already in~\eqref{eq:equation_for_Mb_relu} in the case of the ReLU non-linearity.

	Instead if the non-linearity is bounded, the right hand side of~\eqref{eq:equation_for_Mb} is a bounded function of $\overline{M}$ as well, therefore above a certain critical value of $\theta_s$ it will not possible to find the corresponding value of $\overline{M}$. 
	
	Moreover, if the non-linearity diverges linearly for large arguments, we expect to recover back the free energy and the expressions for the one-layer neuron model for large $\theta_s$. Indeed expanding equation~\eqref{eq:equation_for_Mb} one finds $\overline{M} \sim \theta_s/f_{\text{in}}$ and the effective order parameters reduce to those ones of the perceptron, see~\eqref{eq::effective_order_parameters_perceptron}.

	\subsection{Critical capacity}
	
	In this section we show how in our formalism, it is possible to compute the maximal number of inputs that the neuron is able to classify. We underline that this can be done for a generic form of dendritic non-linearity. 
	
	In the critical capacity limit, the set of possible synaptic weights shrinks towards a single point and $q$ tends to $Q$:
	\begin{equation}
		q = Q - dq.
	\end{equation}
	Correspondingly, the other order parameters scale as
	\begin{subequations}
		\label{eq:scaling_alphac}
		\begin{align}
			\hat q\,, \hat Q &\sim \frac{C}{dq^2}\\
			\hat q - 2\hat Q &\sim \frac{A}{dq} \\
			\hat M & \sim -\frac{B \sqrt{C}}{dq}
		\end{align}
	\end{subequations}
	Using the identities (where $a$ is a positive constant)
	\begin{subequations}
		\begin{align}
			\frac{\int_{0}^{\infty} dx \, x \, e^{-a \frac{x^2}{2} +  b x}}{\int_{0}^{\infty} dx \, e^{-a \frac{x^2}{2} +  b x}} &= \frac{b}{a} + \frac{1}{\sqrt{a}}\frac{G\left(- \frac{b}{\sqrt{a}}\right)}{H\left(- \frac{b}{\sqrt{a}}\right)} \\
			\frac{\int_{0}^{\infty} dx \, x^2 \, e^{-a \frac{x^2}{2} +  b x}}{\int_{0}^{\infty} dx \, e^{-a \frac{x^2}{2} +  b x}} &= \frac{1}{a} + \frac{b^2}{a^2} + \frac{b}{a^{3/2}} \frac{G\left(- \frac{b}{\sqrt{a}}\right)}{H\left(- \frac{b}{\sqrt{a}}\right)}
		\end{align}
	\end{subequations}
	and the expansion
	\begin{equation}
		\label{eq:GHexpansion}
		\frac{G(x)}{H(x)} \simeq x \theta(x)\,, \qquad\text{for} \;|x| \gg 1 \,,
	\end{equation}
	the saddle point equations~(\ref{eq:sp_0}) can be written as
	\begin{subequations}\label{eqs:finalsp2}
		\begin{align}
			\overline{W} &= \frac{\sqrt{C}}{A} \left[ G(B) - B H(B) \right]\\
			Q &= \frac{C}{A^2} \left[ \left( 1+B^2 \right) H(B) - BG(B) \right]\\
			A &= H(B). 
		\end{align}
	\end{subequations}
	More work is required to derive the asymptotic limit of equations~(\ref{eq:sp_1}). First of all we need the expansion of the effective order parameters
	\begin{subequations}
		\begin{align}
			D_0 &= \Delta_q - \Delta_0 = \Delta_Q - \Delta_0 - D_1 \simeq \Gamma_0 - \Gamma_1 dq\\
			D_1 &= \Delta_Q - \Delta_q = \Gamma_1 dq + O(dq^2)
		\end{align}
	\end{subequations}
	where we have defined
	\begin{subequations}
		\begin{align}
			\Gamma_0 &= \Delta_Q - \Delta_0 = \int Dx \, g^2\left(\sqrt{f_{\text{in}}(1-f_{\text{in}}) Q} \, x + f_{\text{in}} \overline M  \right) - \left[ \int D y \, g\left(\sqrt{f_{\text{in}}(1-f_{\text{in}}) Q} \, y + f_{\text{in}} \overline M   \right)\right]^2\\
			\Gamma_1 &= f_{\text{in}}(1-f_{\text{in}}) \int Dz \, \left[ g' \left(\sqrt{f_{\text{in}}(1-f_{\text{in}}) Q} \, z + f_{\text{in}} \overline M \right) \right]^2
		\end{align}
	\end{subequations}
	Now we subtract~\eqref{eq:sp5} with~\eqref{eq:sp6} getting
	\begin{equation}
		\begin{split}
			\hat q - 2 \hat Q &= \alpha \mathbb{E}_\sigma \int Dz \, \frac{G(a_\sigma (z))}{H(a_\sigma(z))} \left[ \frac{z}{\sqrt{D_0 D_1}} \left( \frac{d D_{0}}{d q} +  \frac{d D_{0}}{d Q}\right) -\frac{a_\sigma(z)}{D_1} \left(\frac{d D_1}{dq} +  \frac{d D_1}{dQ} \right) \right] \\
			&= \alpha \mathbb{E}_\sigma \int Dz \, \frac{G(a_\sigma (z))}{H(a_\sigma(z))} \left[ \frac{z}{\sqrt{D_0 D_1}} \frac{d \Gamma_{0}}{d Q}  -\frac{a_\sigma(z)}{D_1} \frac{d \Gamma_1}{dQ} dq\right]
		\end{split}
	\end{equation}
	Similarly~\eqref{eq:sp6} becomes
	\begin{equation}
		\begin{split}
			\hat q &= \alpha \mathbb{E}_\sigma \int Dz \, \frac{G(a_\sigma (z))}{H(a_\sigma(z))} \left[ -\frac{a_\sigma(z)}{D_1} \frac{d D_1}{dq} \right]
		\end{split}
	\end{equation}
	because the second term in~\eqref{eq:sp6} is subleading in $dq$. 
	Using the expansion~(\ref{eq:GHexpansion}) and the identities
	\begin{subequations}
		\begin{align}
			\int Dz \, z^2 \, \Theta(z-\tau_\sigma) &= H(\tau_\sigma) + \tau_\sigma G(\tau_\sigma) \\
			\int Dz \, z \, \Theta(z-\tau_\sigma) &= G(\tau_\sigma)
		\end{align}
	\end{subequations}
	we obtain the following saddle point equations
	\begin{subequations}
		\begin{align}
			0 &= \mathbb{E}_\sigma \sigma \left[ G(\tau_\sigma) - \tau_\sigma H(\tau_\sigma) \right] \\
			A &= \alpha_c \left[ \frac{1}{\Gamma_1 } \frac{d\Gamma_0}{d Q}  \, \mathbb{E}_\sigma H(\tau_\sigma) - \frac{\Gamma_0}{\Gamma_1^2}\frac{d\Gamma_1}{dQ}\mathbb{E}_\sigma\left[ (1+\tau_\sigma^2) H(\tau_\sigma) - \tau_\sigma G(\tau_\sigma)  \right]\right] \\
			C &= \frac{\alpha_c \Gamma_0}{\Gamma_1} \mathbb{E}_\sigma \left[ (1+\tau_\sigma^2) H(\tau_\sigma) - \tau_\sigma G(\tau_\sigma) \right] \,,
		\end{align}
	\end{subequations}
	which involve the critical capacity as an unknown parameter to find. Notice that in the previous equation we have redefined $\tau_\sigma = (\sigma \Delta -\kappa )/ \sqrt{\Gamma_0}$. The full set of saddle point equations for the order parameters $A$, $B$, $C$, $M$, $Q$, $\overline{M}$ and for $\alpha_c$ are
	\begin{subequations}
		\label{eq:sp_alphac}
		\begin{align}
			\overline{W} &= \frac{\sqrt{C}}{A} \left[ G(B) - B H(B) \right]\\
			Q &= \frac{C}{A^2} \left[ \left( 1+B^2 \right) H(B) - BG(B) \right]\\
			A &= H(B) \\
			0 &= \mathbb{E}_\sigma \sigma \left[ G(\tau_\sigma) - \tau_\sigma H(\tau_\sigma) \right] \\
			A &=  \frac{\alpha_c}{\Gamma_1 } \frac{d\Gamma_0}{d Q}  \, \mathbb{E}_\sigma H(\tau_\sigma) - \frac{1}{\Gamma_1}\frac{d\Gamma_1}{dQ} C \\
			C &= \frac{\alpha_c \Gamma_0}{\Gamma_1} \mathbb{E}_\sigma \left[ (1+\tau_\sigma^2) H(\tau_\sigma) - \tau_\sigma G(\tau_\sigma) \right] \\
			\theta_{s} &= \int Dy \, g\left( \sqrt{f_{\text{in}}(1-f_{\text{in}})Q}\,y + f_{\text{in}}\overline{M}\right) \,.
		\end{align}
	\end{subequations}

	\begin{figure*}[]
		\subfloat[ReLU. ]{\includegraphics[width=0.5\columnwidth]{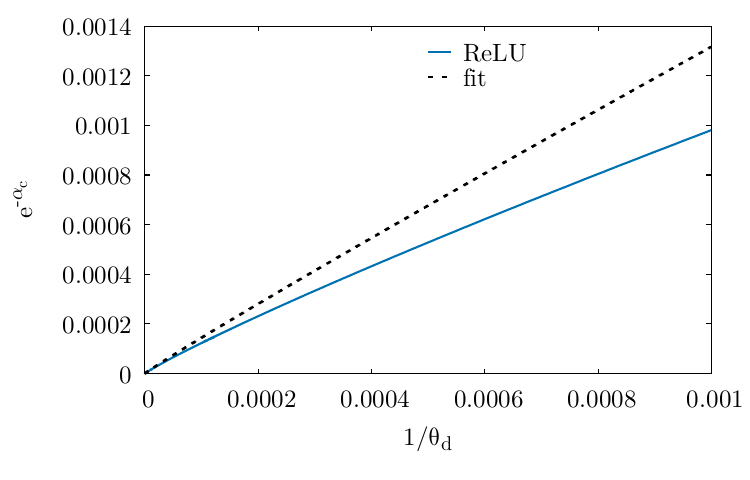}}
		\subfloat[ReLU-Sat.]{\includegraphics[width=0.5\columnwidth]{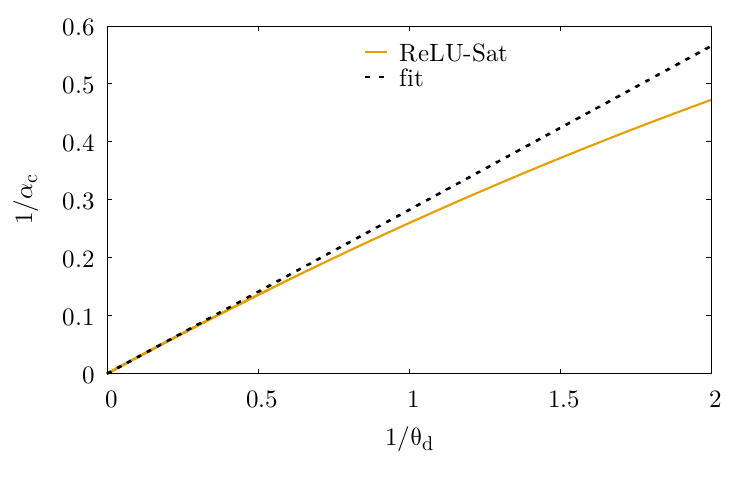} }
		\caption{Fit to the critical capacity for large dendritic thresholds for the ReLU (left panel) and the ReLU-Sat (right panel) non-linearities. The external parameters are the same used in the corresponding figures of the main text, i.e. $f_{\text{in}} = f_{\text{out}} = \theta_s = 0.5$ and $\kappa = 0$. Notice that in the case of the ReLU function (left panel) we are plotting $e^{-\alpha_c}$ versus $\theta_d$ at variance to the ReLU-Sat. In each plot the dashed black line represents a linear fit $a + b x$ of the analytical data for large $\theta_d$. }
		\label{fig:alphac_fit}
	\end{figure*}

	As we have anticipated before, in the case $f_{\text{out}} = 0.5$ the equations can be further simplified, since $M=0$; if also $\kappa=0$ therefore $\tau_\sigma = 0$; the saddle point equations~(\ref{eq:sp_alphac}) then reduce to
	\begin{subequations}
		\begin{align}
			\overline{W} &= \sqrt{\frac{\alpha_c \Gamma_0}{2\Gamma_1}}\frac{1}{H(B)} \left[ G(B) - B H(B) \right]\\
			Q &= \frac{\alpha_c \Gamma_0}{2\Gamma_1} \frac{1}{H^2(B)} \left[ \left( 1+B^2 \right) H(B) - BG(B) \right]\\
			\alpha_c &= \frac{2 \Gamma_{1} H(-B)}{\frac{d\Gamma_0}{d Q}-\frac{\Gamma_0}{\Gamma_1}\frac{d\Gamma_1}{dQ}}\\
			\theta_{s} &= \int Dy \, g\left( \sqrt{f_{\text{in}}(1-f_{\text{in}})Q}\,y + f_{\text{in}}\overline{M}\right) \,.
			\label{eq:alpha_c}
		\end{align}
	\end{subequations}
	
	\subsubsection{Limit of large dendritic threshold}
	
	As mentioned in the main text, the critical capacity of the model depends strongly on the shape of the non-linearity. In particular, in the limit of large dendritic threshold the critical capacity can diverge differently depending if the non-linearity saturates or not for a sufficiently large stimulus. We show in Fig.~\ref{fig:alphac_fit} how the critical capacity diverges logarithmically in $\theta_d$ for the ReLU activation function whereas the divergence is linear in the ReLU-Sat activation function. Performing a fit we get when $\theta_d \to \infty$
	\begin{subequations}
		\begin{align}
			\alpha_c^{\text{ReLU-Sat}} &\simeq 3.518 \, \theta_d \\
			\alpha_c^{\text{ReLU}} &\simeq  \, 0.9602 \, \ln \theta_d
		\end{align}    
	\end{subequations}

	\subsubsection{Limit of small dendritic threshold}
	
	As shown in the main text numerically and above for $\alpha < \alpha_c$, in the low dendritic threshold regime for fixed $\theta_s$ the model with Polsky and ReLU activation function behaves like a one-layer model. We give another quantitative argument here for $\alpha = \alpha_c$. Since $\theta_d \to 0$ the right-hand side of the first of~\eqref{eq:sp_alphac} should go to zero. Since the function $G(B) - B H(B)$ does not go to zero for finite values of $B$, by necessity $C \to 0$. By the last of~\eqref{eq:sp_alphac}, this requires $\Gamma_0 \to 0$, i.e. $Q \to 0$. In this limit we get 
	\begin{equation}
		\Gamma_0 = \Delta_Q - \Delta_0 \simeq \Gamma_1 Q \,,
	\end{equation}
	which holds in the perceptron case. The saddle point equations therefore become equivalent to the one found in the perceptron.

	\subsection{Distribution of synaptic weights}
	The distribution of synaptic weights is:
	\begin{equation}
		P(W) = \Big\langle\Big\langle \frac{1}{\Omega} \int\limits_{0}^{\infty} \prod_{li} dW_{li} \prod\limits_{\mu=1}^{\alpha N} \Theta\left[\frac{\sigma^{\mu}}{\sqrt{K}}\left(\sum\limits_{l=1}^{K}c_{l} \, g\left(\sqrt{\frac{K}{N}}\sum\limits_{i=1}^{N/K} W_{li} \xi^{\mu}_{li}-\sqrt{\frac{N}{K}}\theta{d}\right)-K\theta{s}\right)-\kappa\right]\, \delta (W-W_{11})\Big\rangle\Big\rangle_{\lbrace \xi^{\mu},\sigma^{\mu} \rbrace}
	\end{equation}
	with respect to the first weight of the first sub-perceptron $W_{11}$ for simplicity. Resorting to the replica method by introducing $n$ replicas we have:
	\begin{equation}
		\begin{aligned}
			P(W) =& \lim\limits_{n \to 0} \,\underset{\sigma^{\mu}}{\E} \int\limits_{0}^{\infty} \prod_{i,l,a} dW^{a}_{li} \int\prod_{\mu,a,l} \frac{d\lambda^{a}_{l \mu}\, d\hat{\lambda}^{a}_{l \mu}}{2\pi}\, e^{i\lambda^{a}_{l\mu}\hat{\lambda}^{a}_{l\mu}} \prod\limits_{\mu =1}^{\alpha N} \Theta \left[\frac{\sigma^{\mu}}{\sqrt{K}}\left(\sum\limits_{l} c_{l}\, g(\lambda^{a}_{l\mu})-K\theta{s}\right)-\kappa\right]\, \delta (W-W_{11})\\
			&\times e^{i \sqrt{\frac{N}{K}}\theta{d}\sum\limits_{\mu,a,l}\hat{\lambda}^{a}_{l\mu}}\prod_{l,i,\mu} \Big\langle e^{-i \xi^{\mu}_{li} \sqrt{\frac{K}{N}}\sum\limits_{a} W^{a}_{li}\hat{\lambda}^{a}_{l\mu}}\Big\rangle_{\xi^{\mu}_{li}}
		\end{aligned}
	\end{equation} 
	Repeating the same steps as in section~\ref{sec::replica_method} we have
	\begin{equation}
		\begin{aligned}
			P(W)=& \lim\limits_{n \to 0} \int \prod_{a<b,l} \frac{dq^{ab}_{l}\, d\hat{q}^{ab}_{l}}{2\pi K/N}\int \prod_{a,l} \frac{dQ^{a}_{l}\, d\hat{Q}^{a}_{l}}{2\pi K/N}\int \prod_{a,l} \frac{dM^{a}_{l}\,d\hat{M}^{a}_{l}}{2\pi\sqrt{K/N}}\, e^{-\frac{N}{K}\sum\limits_{a<b,l}q^{ab}_{l}\hat{q}^{ab}_{l} \,-\,\frac{N}{K}\sum\limits_{a,l}Q^{a}_{l}\hat{Q}^{a}_{l}\,-\,\frac{N}{K}\overline{W}\sum\limits_{a,l} \hat{M}^{a}_{l}}\\
			&\times e^{N \alpha \,G_{E} (q^{ab}_{l},Q^{a}_{l},M^{a}_{l}) +(\frac{N}{K}-1)\,G_{S}(\hat{q}^{ab}_{l},\hat{Q}^{a}_{l},\hat{M}^{a}_{l})}\int\limits_{0}^{\infty} \prod\limits_{l,a} dW^{a}_{l} \, \delta(W-W_{1}) \,e^{\sum\limits_{a<b,l}\hat{q}^{ab}_{l}W^{a}_{l}W^{b}_{l}+\sum\limits_{a,l}\hat{Q}^{a}_{l}(W^{a}_{l})^{2}+\sum\limits_{a,l}\hat{M}^{a}_{l} W^{a}_{l}}
		\end{aligned}
	\end{equation}
	where the entropic and energetic terms are the same as in Eqs.~(\ref{eq:Gs}), (\ref{eq:Ge}) and the order parameters are those defined in \eqref{eq:ordpar_comm}. In the limit $n\to 0$ therefore
	\begin{equation}\label{eq:pw}
		P(W) = \lim\limits_{n \to 0}\int\limits_{0}^{\infty} \prod\limits_{l,a} dW^{a}_{l} \, \delta(W-W_{1}) \,e^{\sum\limits_{a<b,l}\hat{q}^{ab}_{l}W^{a}_{l}W^{b}_{l}+\sum\limits_{a,l}\hat{Q}^{a}_{l}(W^{a}_{l})^{2}+\sum\limits_{a,l}\hat{M}^{a}_{l} W^{a}_{l}}
	\end{equation}
	provided the order parameters satisfy the same saddle point equations as before. Under the RS ansatz expression~\eqref{eq:pw} becomes:
	\begin{equation}
		\label{eq:pw_final}
		\begin{split}
			P(W) &= \Theta (W)  \int Dz \, \frac{e^{-\frac{1}{2} (\hat q - 2 \hat Q) W^2 + (\sqrt{\hat{q}} z + \hat M) W }}{\int_0^{\infty} dW \, e^{-\frac{1}{2} (\hat q - 2 \hat Q) W^2 + (\sqrt{\hat{q}} z + \hat M) W }} \\
			&= \Theta (W) \sqrt{\hat{q}-2\hat{Q}} \, e^{-\frac{1}{2}\left(\hat{q}-2\hat{Q} \right)W^2 + \hat{M}W } \int Dz \, e^{\sqrt{\hat{q}} W z} \, \frac{G \left(-\frac{\sqrt{\hat{q}}z+\hat{M}}{\sqrt{\hat{q}-2\hat{Q}}} \right)}{H \left(-\frac{\sqrt{\hat{q}}z+\hat{M}}{\sqrt{\hat{q}-2\hat{Q}}} \right)} 
		\end{split}
	\end{equation}
	Notice that the dependence of $P(W)$ on $K$ and on the activation function is not explicit, but is concealed inside the order parameters that clearly depend on them through the saddle point they have to satisfy. Notice also that for $\alpha = 0$ the synaptic weight satisfies an exponential distribution
	\begin{equation}
		P(W) = \Theta(W) \hat M e^{-\hat M W} = \Theta(W) \frac{f_{\text{in}}}{\theta_d} e^{- \frac{f_{\text{in}}}{\theta_d} W}
	\end{equation}
	This is to be expected, since at $\alpha=0$ the only constrain that is required apart for the fact that the synapses are non-negative, is that their average is $\overline{W} = \frac{\theta_d}{f_{\text{in}}}$.

	\subsubsection{Distribution of synaptic weights in the maximal storage limit}
	In the critical capacity limit $\alpha \to \alpha_c$ the expression of the distribution of synaptic weight greatly simplifies. Using the scalings in~\eqref{eq:scaling_alphac} we find
	\begin{equation}
		P(W) = \Theta (W) \, e^{-\frac{A}{2dq}W^2 - \frac{B\sqrt{C}}{dq}W } \sqrt{\frac{A}{dq}}\int Dz \, e^{\frac{\sqrt{C}}{dq}Wz}  \left[ G \left(\sqrt{\frac{C}{A}}\frac{z-B}{\sqrt{\hat{q}-2\hat{Q}}} \right) \Theta(z-B) - \sqrt{\frac{C}{A dq}} (z-B)\Theta(B-z)\right] 
	\end{equation}
	Using the identity
	\begin{equation}
		\int Dz \, e^{az} \,(z+b) \, \Theta(-b - z) = (a+b) \, e^{\frac{a^2}{2}} H(a+b) - e^{-ab} \, G(b)
	\end{equation}
	we obtain
	\begin{equation}
		P(W) = H(-B) \, \delta(W) + \frac{1}{\sqrt{2\pi} W_\star} e^{-\frac{(W + B W_\star)^2}{2W_\star^2}} \Theta (W) 
	\end{equation}
	where $W_\star \equiv \frac{\sqrt{C}}{A}$. As showed in~\cite{Brunel2004} in the one layer neuron model the synaptic weight distribution changes from being exponential to being a Gaussian plus a spike consisting to a fraction $H(-B)$ of ``silent'' weights at the critical capacity. Indeed as constraints due to the training set are added, more and more synapses tend to assume low weight. This is the case also in our two layer neuron model. It is interesting to note that both the distribution of synaptic weight at finite $\alpha$~\eqref{eq:pw_final} and at critical capacity are in form exactly the same as the one derived in~\cite{Brunel2004} for the one layer neuron model; the dependence on the non-linearity induced by the dendrites is actually implicit in the order parameters. 
	
	\begin{figure*}[h]
		\subfloat[\label{fig:silent_vs_xmin} $\kappa=0$, $f_{\text{in}} = f_{\text{out}} = 0.5$ and $\theta_s = \theta_d = 0.5$, $\gamma = 15$. ]{\includegraphics[width=0.5\columnwidth]{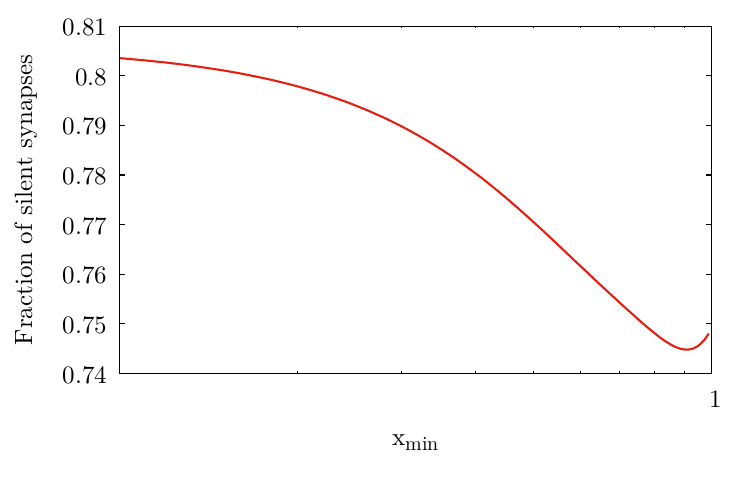}}
		\subfloat[\label{fig:silent_vs_gamma}$\kappa=0$, $f_{\text{in}} = f_{\text{out}} = 0.5$ and $\theta_s = \theta_d = 0.5$, $x_{\text{min}} = 0.33$.]{\includegraphics[width=0.5\columnwidth]{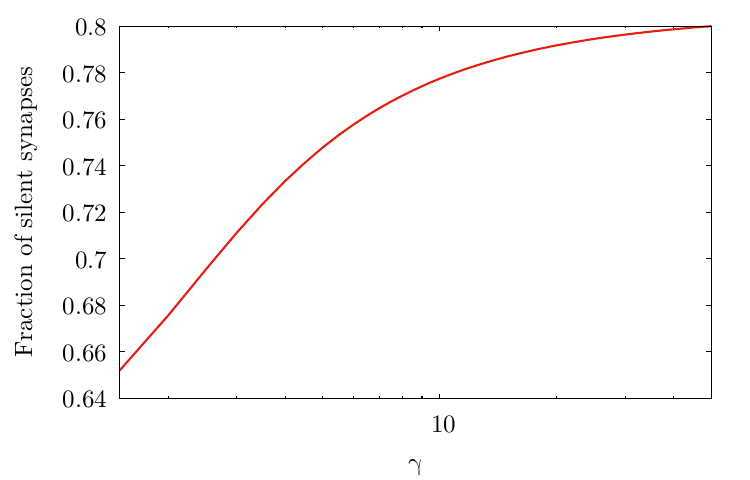} }
		\hspace{1cm}
		\centering
		\subfloat[\label{fig:silent_vs_thetas}$\kappa=0$, $f_{\text{in}} = f_{\text{out}} = 0.5$ and $\theta_d = 0.5$, $x_{\text{min}} = 0.33$, $\gamma = 15$.]{\includegraphics[width=0.5\columnwidth]{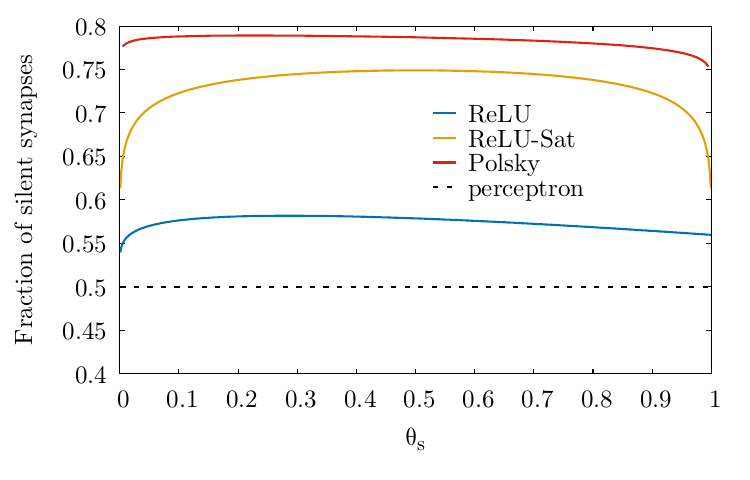} }
		\caption{In the top panels we plot the fraction of silent synapses for the Polsky non-linearity as a function the parameters $x_{\text{min}}$ and $\gamma$. The bottom panel shows the fraction of silent synapses as a function of the somatic threshold, comparing the ReLU, the ``saturating'' ReLU and Polsky non-linearity when no robustness parameter $\kappa$ is imposed. The dashed black line represents the case of the one-layer neuron model, where the critical capacity $\alpha_c^{\text{perc}} = 1$. In the captions of the panels we show the value of the fixed external parameters. }
		\label{fig:silent_vs_thetas}
	\end{figure*}
	
	In Fig.~\ref{fig:alpha_c} we show how the fraction of silent synapses $p_0 = H(-B)$ depends on the somatic threshold for the ReLU, ReLU-Sat and Polsky non-linearities. We also show in the Polsky case, how $p_0$ depends on the parameters defining the shape of the function itself, $x_{\text{min}}$ and $\gamma$.

	\section{Choice and scaling of the numerical simulations hyper-parameters}
	
	Consider the transfer function implemented by the neuron with non-linear dendritic branches, which represents the output preactivation prior to the thresholding operation performed by the $\Theta$-function:
	\begin{equation}
		\label{eq:neuron_supp2}
		\Delta^{\mu}_{\text{out}}
		= \frac{1}{\sqrt{K}}\sum_{l=1}^{K} c_{l} \, g\left(\sqrt{\frac{K}{N}} \sum_{i=1}^{N/K} W_{li} \xi_{li}^\mu - \sqrt{\frac{N}{K}}\theta_{d}\right) - \sqrt{K} \theta_s
	\end{equation}
	we impose that $W \in \left[0, \frac{2\theta_d}{f_{in}}\right]$ and consequently:
	$$
	W \sim O\left(\frac{\theta_d}{f_{in}}\right).
	$$
	If we assume that:
	$
	\theta_d \sim O(1),
	$ and 
	$
	\theta_s \sim O(1)
	$
	we have:
	$
	W \sim O\left(\frac{1}{f_{in}}\right).
	$
	
	We also have:
	$ 
	\sum_{i=1}^{N/K} W_{li} \xi_{li}^\mu  \sim O\left(\frac{N}{K}\right).
	$
	Consequently, from \eqref{eq:neuron_supp2}, we observe that the dendritic pre-activations scale as:
	\begin{equation}
		\left(\sqrt{\frac{K}{N}} \sum_{i=1}^{N/K} W_{li} \xi_{li}^\mu - \sqrt{\frac{N}{K}}\theta_{d}\right) \sim O\left(\sqrt{\frac{N}{K}}\right).
	\end{equation}
	Consequently, the pre-activation variance remains finite in the limit $N\to \infty$, which is the primary motivation for choosing these scalings, as the pre-activation variance is a crucial factor in ensuring the consistency of the learning setting when varying the input dimensionality $N$ and the number of dendritic branches $K$.
	
	Applying the same type of consideration to the pre-activation of the single output node in \eqref{eq:neuron_supp2} and recalling that $c_l=1 \,\, \forall \,l$ and $g(\cdot)\sim O(1)$, we find that it scales as:
	\begin{equation}
		\Delta^{\mu}_{\text{out}} \sim O\left(\theta_s\sqrt{K}\right)
		\label{eq:preact_scaling}
	\end{equation}

	\paragraph{Stochastic Gradient Descent (SGD) with cross-entropy loss}
	
	Recalling the expression for the cross-entropy loss used to investigate the performance of stochastic gradient descent (SGD) on the neuron:
	\begin{align}
		\mathcal{L}_{ce}\left(\Delta^{\mu}_{\text{out}}\right) = \frac{1}{2\gamma_{ce}} \log \left( 1 + \exp\left(-2\gamma_{ce}\Delta^{\mu}_{\text{out}}\right)\right)
	\end{align}
	we can estimate the gradients of the cross-entropy loss, which are used in the SGD update, as follows:
	\begin{eqnarray}
		\frac{\partial \mathcal{L}_{ce}\left(\Delta\right)}{\partial \Delta}  
		&= - \frac{1}{1 + \exp\left(2\gamma_{ce}\Delta\right)}\\
		&\sim O(1)
		\label{eq:scaling_loss_dev}
	\end{eqnarray}
	where the last step holds (due to \eqref{eq:preact_scaling}) if:
	\begin{equation}
		\gamma_{ce} \sim O\left(\frac{1}{\theta_s\sqrt{K}} \right).
	\end{equation}
	
	Proceeding with the derivative w.r.t. to the weights, we obtain:
	\begin{align}
		\frac{\partial \mathcal{L}_{ce}\left(\Delta(W)\right)}{\partial W}  
		&= \frac{\partial \Delta(W)}{\partial W} \frac{\partial \mathcal{L}_{ce}\left(\Delta\right)}{\partial \Delta} \\
		&= \frac{\partial}{\partial W} \left( \frac{1}{\sqrt{K}}\sum_{l=1}^{K} c_{l} \, g\left(\sqrt{\frac{K}{N}} \sum_{i=1}^{N/K} W_{li} \xi_{li}^\mu - \sqrt{\frac{N}{K}}\theta_{d}\right) - \sqrt{K} \theta_s \right) \frac{\partial \mathcal{L}_{ce}\left(\Delta\right)}{\partial \Delta} \\
		&= \left[ \frac{1}{\sqrt{K}}\sum_{l=1}^{K} \left( c_{l} \, g'\left(\sqrt{\frac{K}{N}} \sum_{i=1}^{N/K} W_{li} \xi_{li}^\mu - \sqrt{\frac{N}{K}}\theta_{d}\right) 
		\sqrt{\frac{K}{N}} \sum_{i=1}^{N/K} \xi^{\mu}_{il}
		\right) \right]
		\frac{\partial \mathcal{L}_{ce}\left(\Delta\right)}{\partial \Delta} \\
		&\sim O\left( \frac{1}{\sqrt{K}} \sqrt{\frac{K}{N}} f_{in} N \right) O(1) \\
		&\sim O\left( f_{in} \sqrt{N} \right)
	\end{align}
	where in the penultimate step, we have used the fact that the derivative of the function $g$ is bounded within the interval $[0,1]$ so that $g'(\cdot)\sim O(1)$; that $\sum_{i=1}^{N/K} \xi^{\mu}_{il} \sim O\left(f_{in}\frac{N}{K}\right)$; and \eqref{eq:scaling_loss_dev}.
	
	At this point, recalling that the SGD update rule is:
	\begin{equation}
		w_{il} \leftarrow w_{il} - \zeta \nabla_{w_{il}} \mathcal{L}(w_{il})
	\end{equation}
	and imposing an update of the same order as the weights, i.e., $w\sim O(\theta_d) \implies \zeta \nabla_w \mathcal{L}(w) \sim O(\theta_d)$, we find the desired scaling for the learning rate $\zeta$:
	\begin{equation}
		\zeta \sim O\left( \frac{\theta_d}{f_{in}\sqrt{N}} \right)
	\end{equation}
	
	\paragraph{Comparison between the dendritic and the linear neuron}
	
	The transfer function of the linear neuron (i.e. the perceptron model), is given by:
	\begin{equation}
		\label{eq:perceptron_supp}
		\Delta^{\mu}_{\text{perc, out}}
		= \frac{1}{\sqrt{N}} \sum_{i=1}^{N} W_{i} \xi_{i}^\mu - \sqrt{N} \theta_s.
	\end{equation}
	From this expression, and observing that
	the weights scale with $\theta_s$, we obtain the following scalings for the learning rate $\zeta$ and cross-entropy parameter $\gamma_{ce}$:
	\begin{equation}
		\zeta \sim O\left( \frac{\theta_s}{f_{in}\sqrt{N}} \right), \qquad
		\gamma_{ce} \sim O\left( \frac{1}{\theta_s\sqrt{N}} \right).
	\end{equation}

	\twocolumngrid 
	
	\bibliography{references}

\end{document}